\documentclass[draft]{agujournal2019}
\usepackage{url} 
\usepackage{lineno}
\usepackage[inline]{trackchanges} 
\usepackage{soul}
\usepackage{amsmath}
\usepackage{amssymb}
\usepackage{comment}
\draftfalse

\journalname{JGR: Space Physics}

\newcommand{\aap}{{\it Astron. Astrophys.}}

\newcommand{\apj}{{\it Astrophys. J.}}

\newcommand{\jgr}{{\it J. Geophys. Res.}}

\newcommand{\solphys}{{\it Sol. Phys.}}

\begin{document}

\title{Preconditioning of the interplanetary medium due to isolated ICMEs}

\authors{Primo\v{z} Kajdi\v{c}\affil{1}, Manuela Temmer\affil{2}, Xochitl Blanco-Cano\affil{1}}

\affiliation{1}{Space Science Department, Geophysics Institute, Universidad Nacional Autónoma de México, Mexico City, Mexico}
\affiliation{2}{Institute of Physics, University of Graz, Universitätsplatz 5/II, A-8010 Graz, Austria}

\correspondingauthor{Primo\v{z} Kajdi\v{c}}{primoz@igeofisica.unam.mx}

\begin{keypoints}
\item Statistical study quantifying the potential preconditioning of the interplanetary medium by isolated interplanetary coronal mass ejections (ICMEs).
\item Behind ICMEs, on average the solar wind is less dense, faster, and the magnetic field is stronger and more radial than upstream.
\item Our results suggest that slow ICMEs precondition the IMF more than fast ones.

\end{keypoints}

\begin{abstract}
We perform a systematic study of the preconditioning of the interplanetary (IP) medium due to isolated interplanetary coronal mass ejections (ICMEs). Preconditioning is highly relevant when ICMEs, ejected in close succession and direction, modify the IP medium to allow subsequent ICMEs to propagate more freely, decelerate less, and retain higher kinetic energy at larger distances. 
We base our study on a sample of carefully selected events. The IP medium is analyzed during time intervals of 48 hours before and after the ICMEs in order to statistically quantify their impact on the properties of the solar wind (SW) and interplanetary magnetic field (IMF). We find that the SW behind ICMEs on average exhibits reduced density ($-$41\%) and dynamic pressure ($-$29\%), and increased total velocity (+15\%), while the trailing IMF is more intense (+14\%) and more radially aligned (13$^\circ$). The results suggest that even relatively low speed ICMEs can significantly precondition the IP medium. The results are relevant for better understanding of CME propagation and SW interaction, and hold implications for heliospheric models and applied research of space weather.
\end{abstract}

\section*{Plain Language Summary}
Interplanetary coronal mass ejections (ICME) have been recognized as the drivers of the most intense geomagnetic storms. This study looks at how isolated ICMEs change the space environment around them, making it easier for later ICMEs to travel. When ICMEs are ejected close together and in the same direction, they can reduce resistance in the solar wind, allowing the following ICMEs to move faster and retain more energy. By analyzing selected events and studying the solar wind (SW) and magnetic field 48 hours before and after each ICME, we found that the plasma behind an ICME tends to have lower density and pressure but higher speed and magnetic field strength. This means that even smaller ICMEs can significantly alter their surroundings. These findings help improve our understanding of SW behavior and space weather predictions.

\section{Introduction}
\label{introduction}
The solar wind (SW) is a flow of plasma continuously emitted by the Sun that propagates radially towards the edge of the heliosphere and permeates all of the interplanetary (IP) space. With it, the SW carries the interplanetary magnetic field (IMF) that also originates at the Sun. The constant outflow of SW is occasionally disturbed by transient solar ejecta, such as interplanetary coronal mass ejections (ICMEs). Fast ICMEs have been known to drive large-scale compression waves that may steepen into collisionless IP shocks \cite{bame:1979, tsurutani:1988, schwenn:2006}. About a third of those ICMEs that drive IP shocks exhibit in-situ observational signatures of magnetic clouds, namely smooth rotations of the magnetic field vector, increased magnetic field magnitude, low plasma beta (ratio between thermal and magnetic pressures, e.g. $\beta = nk_{\rm B}T/\frac{B^2}{2\mu_{\rm 0}}$), low densities and bidirectional suprathermal strahl electrons \cite{burlaga:1981, burlaga:1991}. These signatures led \citeA{marubashi:1986} to infer that magnetic clouds exhibit a flux rope topology. In the meantime, it is acknowledged that flux rope structures might be an intrinsic property of every ICME \cite{vourlidas:2013}. To have a more general definition of the magnetic structure of an ICME not necessarily following all criteria for a magnetic cloud, we use in the following the term magnetic obstacle (MO) instead of magnetic cloud.

It has been known for a long time that ICMEs, especially the fast ones, modify the properties of the IP medium, i.e. the SW and the IMF. For example, the ejecta themselves distort the IMF mainly due to the field line draping \cite{mccomas:1989}. On the other hand, the IP shocks compress, accelerate (in the Earth reference frame), deflect and heat the ambient solar wind creating a shock-sheath region between the shock and the ejecta \cite<e.g.,>[]{kilpua:2017a}. Due to increased SW velocity, density and B-magnitude inside the sheaths and/or MO, the ICMEs are the drivers of the most intense geomagnetic storms \cite{tsurutani:1988, mccomas:1989}. 

The dynamics of the ICMEs is governed by three major forces: the Lorentz force ($F_{\rm L}$), the gravitational force ($F_{\rm g}$) and the drag force ($F_{\rm D}$) \cite{vrsnak:2001b}. The latter appears as a consequence of the interaction of the ICMEs with the SW. The drag force may be expressed as
\begin{equation}
F_D = -\rho_{e} A C_D \large(V_i - V_e\large)\lvert V_i - V_e\rvert,
\label{eq:drag}
\end{equation}
where A is the cross-sectional area of the ICME, $C_{\rm D}$ is the drag coefficient and the subscripts $i$ and $e$ refer to quantities inside the ICME and external to the ICME (solar wind) \cite<i.e.,>[]{cargill:2004}. The minus sign means that when the ICME is faster than the preceding SW, the $F_{\rm D}$ points sunward. Close to the Sun, the dynamics of ICMEs is predominantly governed by the Lorentz force. Farther in the IP space, the drag force becomes dominant. 

\citeA{cargill:2004} analyzed the ICME propagation based on the equation~\ref{eq:drag} and showed that higher initial $\rho_{\rm i}/\rho_{\rm e}$ ratios favor lower deceleration rates for ICMEs propagating faster than the preceding SW. \citeA{vrsnak:2010} also studied solutions of a drag-based equation of motion and concluded that the ICME speeds are influenced by their mass and shape and the properties of the SW through which they propagate. The shortest transit times and highest velocities at 1~A.U. were shown to be linked to dense and narrow ICMEs,
which propagate through regions of high-speed, and hence lower-density, SW streams. The term ``narrow'' refers to small CME-cone angular width used to describe the ejecta in the CME-cone model \cite<see, for example>{fisher:1984}. Smaller cone angle means lower surface area of the CME which then translates to smaller drag acceleration.

The drag force decelerates those ICMEs that are faster than the preceding SW and accelerates the slower ones \cite{cargill:1996, gopalswamy:2000, vrsnak:2001}. It is commonly represented in its aerodynamic form in numerous studies and it considers only SW parameters without incorporating the IMF. However, some past studies indicate that the IMF orientation also plays a role in ICME propagation. For example, \citeA{liu:2014}, suggested that radially oriented IMF favors higher ICME propagation speeds. However, this effect has not been well explored, which is a clear shortcoming. Therefore, this study also seeks to advance the understanding of ICME-SW interactions by integrating IMF information into the analysis.

The interaction between ICMEs and SW/IMF is one of the main reasons why huge eruptive events that show the highest velocities near the Sun do not always lead to the most intense space weather events on Earth. In order to cause the most extreme geomagnetic storms, the ICMEs must carry a strong negative $B_z$  during an extended interval, and be fast at 1~A.U. Favorable IP conditions for ICMEs to maintain their high propagation speeds can be achieved through an effect known as preconditioning \cite{desai:2020, liu:2019}.

This occurs when one or more ICMEs are ejected in roughly the same direction in the course of a few days. These ``clear'' the IP space, leaving behind a wake with a lower density and faster SW and more radially oriented IMF in which any subsequent ICME can propagate much more freely. Since it involves two or more events, \citeA{liu:2019} described preconditioning as an indirect interaction between ICMEs. These authors also suggested that preconditioning could be the only way to produce ejecta with magnetic field magnitude above 100~nT and a SW transient speed higher than 2000 kms$^{-1}$ at 1 A.U.

In fact, preconditioning has been invoked several times in the past in order to explain some of the shortest ICME travel times and the most intense geomagnetic storms in the history, including the succession of superstorms in January 1938 \cite{hayakawa:2021}, the August 4 1972 event, which caused unintended detonation of dozens of sea mines in North Vietnam \cite{knipp:2018}, the 13 March 1989 event famous for causing a blackout of the Hydro-Québec system \cite{boteler:2019} and the fastest ever ICME detected in-situ on 24 July 2012 \cite{liu:2014,temmer:2015}.

To our knowledge, there exist only a few studies that deal with the preconditioning of the SW in a systematic manner. \citeA{temmer:2017} quantified the duration of disturbed conditions in the IP space caused by ICMEs. These authors studied the speed of the SW impacted by ICMEs during the period 2011-2015 and compared it to different background solar wind models, namely the empirical solar wind forecast model, the Wang–Sheeley–Arge model and the 27-day persistence model  \cite<a comparison between these models is presented e.g., in>[]{reiss:2016}. The results by \citeA{temmer:2017} showed that the periods within the ICMEs exhibited a speed increase of 18~\%--32~\% above the expected background, while the periods two days after the ICMEs displayed an increase of 9~\%--24~\%. The total duration of the enhanced speeds lasted for up to five days after the ICME start, which is much longer than the average duration of an ICME disturbance itself ($\sim$1.3 days).

\citeA{wu:2022} performed numerical simulations in which they compare the propagation of ICMEs and their shocks that were preceded by pre-events to those that were not. The authors found that in the case that pre-events existed, the ICME shocks exhibited higher speeds. It was concluded that the shocks accelerated once they entered regions modified by the pre-events inside which the SW was characterized by lower densities, lower temperatures and higher speeds.

In this work, we take a different, fully observational approach, focusing only on the IMF and SW properties before the ICME arrival and after their passage. Focusing on ICME-SW interaction, we do that for strictly isolated ICMEs, i.e., not preceded or followed by another large-scale structure (CME, CIR) in the IP medium. For that purpose we carefully selected a sample of events for which we examine how the SW plasma and IMF properties change between the periods before and after the ICMEs.

Due to our strict requirements, our sample consists of 21 ICMEs detected by four different missions. 

This paper is organized as follows: in Section~\ref{sec:data} we present the datasets and explain the procedure through which we selected suitable events. In Section~\ref{sec:results} we present the statistical results of our study and in Section~\ref{sec:discussion} we discuss and summarize our findings. Finally, we present our conclusions in Section~\ref{sec:conclusions}.

\section{Data, Selection of Events and Methods}
\label{sec:data}
In order to search for suitable events, we examined the ICMEs listed in the ``Interplanetary Coronal Mass ejections multi-catalog''\footnote{\url{https://edatos.consorciomadrono.es/dataset.xhtml?persistentId=doi:10.21950/XGUIYX}} compiled by \citeA{larrodera:2024}. The catalog itself covers twelve different ICME catalogs from different missions over the period 1975--2022. The events selected for the purpose of this study were detected by the ACE \cite{stone:1998}, STEREO-A and STEREO-B \cite{kaiser:2008} and Helios-1 \cite{porsche:1981, wenzel:1992} missions.

In order to study the IP medium preconditioning by the ICMEs, we opted to compare the observed SW and IMF properties during 48 hours prior to the ICME arrival and during 48 hours after their passage. We used the ICME start and end times listed in the aforementioned catalog. Our definition of ICMEs include their corresponding IP shocks, shock-sheath regions as well as MOs. The ICMEs in our sample were required to be isolated events. To be classified as isolated event no additional large-scale solar wind structures, such as other ICMEs or co-rotating interaction regions \cite<CIRs;>[]{gosling:1999} had to be detected 48 hours before and after the event. Hence, the ICMEs had to be very well-defined events, with clear boundaries, as opposed to the complex ones, where either two or more ejecta or one ejecta and a CIR merge into a single event. It was also required that the maximum speed inside the ICMEs had to be higher than that of the trailing SW. 

Finally, only events observed close to 1~A.U. were included in the sample.

Out of the hundreds of events in the catalog, only 21 satisfied our criteria (see Table~\ref{tab:1}). Of those, three were detected by the STEREO-A spacecraft, six by STEREO-B, 11 by ACE and one by the Helios-1 mission.

For the purpose of this study we use the magnetic field data from the ACE Magnetic Fields Experiment \cite{smith:1998}, STEREO/IMPACT Magnetic Field Experiment \cite{acuna:2008},
and Helios-1 magnetometer \cite{porsche:1981}. Plasma measurements used in this work were obtained by the Solar Wind Electron Proton Alpha Monitor (SWEPAM) instrument onboard ACE \citeA{mccomas:1998}, Measurements of Particles And CME Transients Suite (IMPACT) onboard the STEREO probes \cite{luhmann:2008}, 
and Helios Plasma Detector \cite{porsche:1981}.

The magnetic field components are presented in RTN coordinate system (R component points radially outward from Sun, T component is approximately aligned with the spacecraft orbital direction while N points northward; the RN plane contains the solar rotation axis) with time resolution of $\lesssim$1 minute. The plasma moments used in this work come in similar range of time resolutions and the SW velocity components are also provided in RTN system. In the case of Helios-1, we use the merged magnetic field and plasma datasets with one hour time resolution provided by the Coordinated Data Analysis Web (CDAWeb, https://cdaweb.gsfc.nasa.gov/).

\begin{center}
\begin{table}
\begin{tabular}{c c c c c c c c c}
\hline
Start date & Start time & End date & End time & Start date & Start time & R$_{heliospheric}$ & Duration & Mission\\
ICME& ICME& ICME& ICME& MO& MO& [AU] & (Days) &\\
\hline
\hline\\
2010-09-17 & 22:34:00 & 2010-09-18 & 06:22:00 & 2010-09-19 & 06:21:00 & 0.97 & 1.32 & STA\\
2013-06-20 & 11:13:00 & 2013-06-20 & 11:13:00 & 2013-06-21 & 09:37:00 & 0.96 & 0.93 & STA\\
2013-08-22 & 07:05:00 & 2013-08-22 & 23:15:00 & 2013-08-24 & 23:25:00 & 0.97 & 2.68 & STA\\
2011-02-26 & 08:28:00 & 2011-02-26 & 16:00:00 & 2011-02-27 & 23:00:00 & 1.02 & 1.61 & STB\\
2011-06-01 & 07:25:00 & 2011-06-01 & 17:35:00 & 2011-06-02 & 18:00:00 & 1.01 & 1.44 & STB\\
2012-07-04 & 06:56:00 & 2012-07-04 & 11:40:00 & 2012-07-05 & 12:50:00 & 1.01 & 1.25 & STB\\
2013-12-08 & 18:22:00 & 2013-12-09 & 02:25:00 & 2013-12-10 & 18:40:00 & 1.08 & 2.01 & STB\\
2014-03-14 & 23:10:00 & 2014-03-15 & 05:15:00 & 2014-03-16 & 18:12:00 & 1.06 & 1.79 & STB\\
2014-04-12 & 02:27:00 & 2014-04-12 & 20:46:00 & 2014-04-15 & 07:13:00 & 1.04 & 3.20 & STB\\
1998-11-30 & 04:17:00 & 1998-11-30 & 09:42:00 & 1998-12-01 & 02:34:00 & 0.97 & 0.93 & ACE\\
1998-08-26 & 06:20:00 & 1998-08-26 & 21:26:00 & 1998-08-28 & 00:39:00 & 1.00 & 1.76 & ACE\\
1999-04-16 & 10:35:00 & 1999-04-16 & 18:00:00 & 1999-04-17 & 19:00:00 & 0.99 & 1.35 & ACE\\
2001-04-28 & 04:31:00 & 2001-04-28 & 15:53:00 & 2001-05-01 & 03:38:00 & 0.99 & 2.96 & ACE\\
2001-08-17 & 10:15:00 & 2001-08-17 & 20:52:00 & 2001-08-19 & 03:27:00 & 1.00 & 1.72 & ACE\\
2003-11-20 & 07:27:00 & 2003-11-20 & 10:06:00 & 2003-11-21 & 00:42:00 & 0.98 & 0.72 & ACE\\
2005-05-15 & 02:10:00 & 2005-05-15 & 05:29:00 & 2005-05-17 & 02:09:00 & 1.00 & 2.00 & ACE\\
2010-04-05 & 07:54:00 & 2010-04-05 & 12:01:00 & 2010-04-06 & 13:20:00 & 0.99 & 1.23 & ACE\\
2010-08-03 & 16:54:00 & 2010-08-04 & 09:41:00 & 2010-08-05 & 00:00:00 & 1.00 & 1.30 & ACE\\
2012-07-14 & 17:26:00 & 2012-07-15 & 05:52:00 & 2012-07-17 & 06:46:00 & 1.00 & 2.56 & ACE\\
2016-07-19 & 23:02:00 & 2016-07-20 & 06:37:00 & 2016-07-22 & 11:52:00 & 1.00 & 2.53 & ACE\\
1977-01-29 & 01:00:00 & 1977-01-29 & 11:02:00 & 1977-01-30 & 06:02:00 & 0.98 & 1.21 & Helios 1\\
\hline
\end{tabular}
\caption{List of isolated ICMEs.}\label{tab:1}
\end{table}
\end{center}

\begin{figure*}
    \centering
    \includegraphics[width=0.7\linewidth]{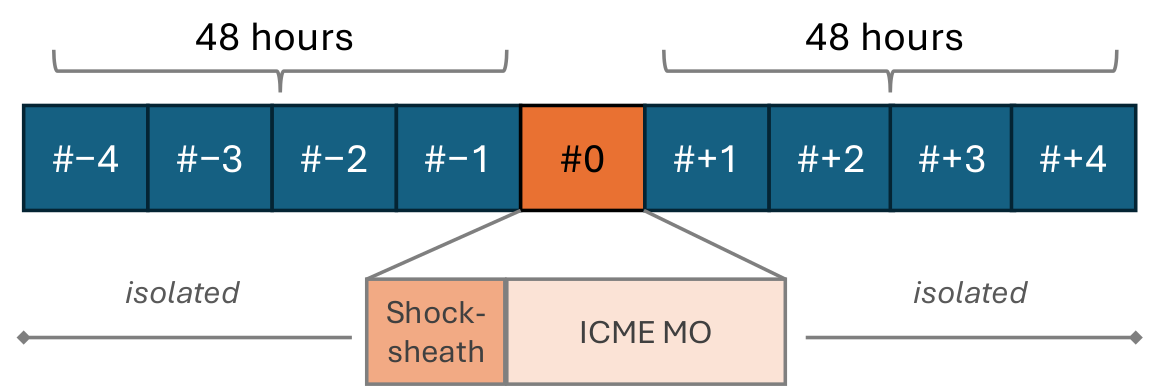}
    \caption{Intervals used. Each box before and after the ICME structure marks a 12 hours region; values for the shock-sheath and ICME MO interval are averaged and represented as one data point (i.e., duration has no meaning).}
    \label{fig:intervals}
\end{figure*}

After the careful manual selection of events, we calculate average values of magnetic field magnitude and cone angle and plasma moments. In addition, we derive the average dynamic pressure $P_{dyn}$ (defined as $\rho v^2$, with $\rho$ the mass density of the plasma, and $v$ the velocity of the plasma), thermal pressure $P_{th}$ (defined as $nk_B T$ with $n$ the number density of particles, $k_B$ the Boltzmann constant, and $T$ the plasma temperature), and magnetic pressure $P_B$ (defined as $\frac{B^2}{2\mu_0}$ with $B$ the IMF magnitude and $\mu_0$ the permeability of free space). We also calculate the average IMF cone angle (defined as $\arccos\left(\frac{B_R}{B}\right)$ with $B$ the IMF magnitude and $B_R$ its radial component). We do this by averaging the corresponding spacecraft measurements during four consecutive 12 hour intervals. For each physical quantity we thus obtain four values prior to the ICME arrivals and four values after the ending of the MOs. The interval marked as \#0 refers to average values derived over the entire period of the ICME including the shock, the sheath region and the MO. Points with negative sign (\#$-$1 to \#$-$4) correspond to four successive 12 hour averages prior to the ICME arrival. Those with positive sign (\#$+$1 to \#$+$4) refer to the four successive 12 hour intervals after the end of the ICME (see Figure~\ref{fig:intervals}). 

\section{Results}
\label{sec:results}
\subsection{Case study}
\begin{figure}
\centering
\includegraphics[width=1.0\textwidth]{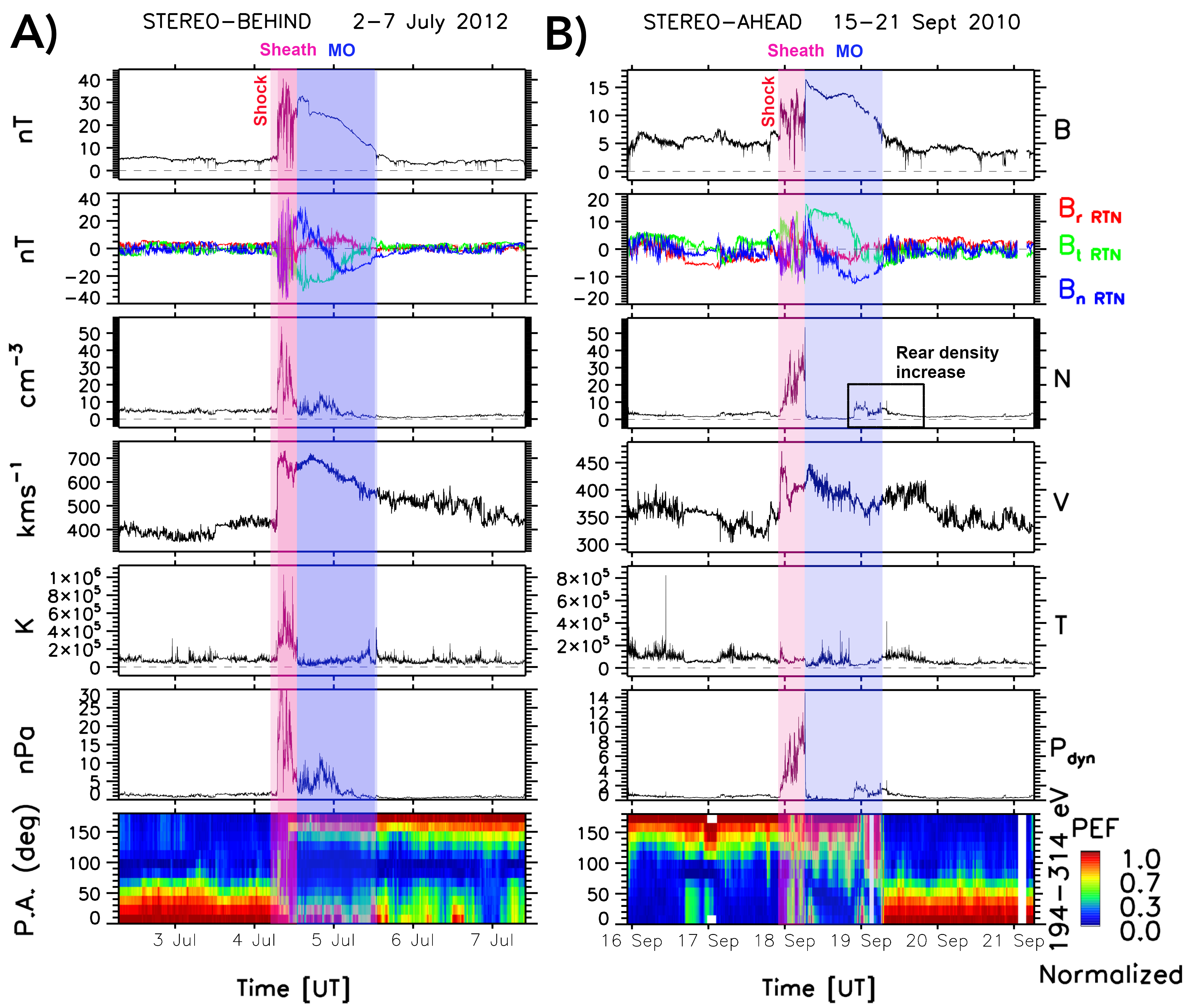}
\caption{Two examples of isolated, magnetic-cloud type ICMEs and their surroundings observed by the STEREO probes. Left: a fast ICME observed during 4--5 July 2012. Right: a slow event observed during 17--19 September 2010. Both ICMEs developed IP shocks and the corresponding shock-sheath regions, that were followed by MOs with magnetic cloud characteristics. Panels show, from top ot bottom: IMF magnitude in units of nanoTesla (nT), magnetic field components in the RTN coordinate system, SW number density (cm$^{-3}$), velocity (kms$^{-1}$), plasma temperature (K), dynamic pressure (nPa) and electron pitch angle spectrogram, where the colors represent the normalized particle energy flux (PEF) in the 194-312 eV energy range, where the PEF at each point in time is normalized to its highest value at that time. The reason for normalizing is to emphasize the fluxes of bidirectional strahl electrons inside MOs.}
\label{fig:icme}
\end{figure}

Two examples of ICMEs and their surroundings from our sample are presented in Figure~\ref{fig:icme}. The ICMEs were observed by the STEREO probes during 4--5 July 2012 (Figure~\ref{fig:icme}A) and 17--19 September 2010 (Figure~\ref{fig:icme}B) and we also exhibit 48-hour intervals before and after the events. Panels show, from top to bottom: IMF magnitude in units of nanoTesla (nT), magnetic field components in the RTN coordinate system (nT), SW number density (cm$^{-3}$), velocity (kms$^{-1}$), plasma temperature (K), dynamic pressure (nPa) and electron pitch angle (PA) spectrogram. The colors in the spectrogram represent the normalized particle energy flux (PEF) in the 194--312 eV energy range. This range was chosen in order to display the PA distributions of suprathermal strahl electron population whose typical energy range is between 70~eV and 2~keV \cite<see for example>[]{kajdic:2016}. The PEF is normalized to its highest value at each each point in time. The reason for normalizing is to emphasize the fluxes of bidirectional strahl electrons that are represented on these panels with red and green shades. The pitch angle on the vertical axis is an angle between the electron velocity and the IMF vectors.

The July 2012 ICME exhibits an IP shock (red vertical line in Figure~\ref{fig:icme}A), sheath region (purple shading) and an MO with a typical magnetic cloud-like signature (blue shading) with enhanced magnetic field magnitude (up to $\lesssim$40~nT), smooth IMF rotations, reduced density and temperature and bidirectional suprathermal strahl electrons. During 48 hours before the shock arrival and 48 hours after the end of the MO, the SW is relatively quiet and does not exhibit any major structures. Thus, this event has been classified as an isolated event.

The September 2010 event (Figure~\ref{fig:icme}B) also drove an IP shock and the MO exhibited properties of a magnetic cloud, such as enhanced B-magnitude, smooth rotations of the magnetic field, diminished density and short, intermittent intervals of bidirectional strahl electrons. However, this event exhibits a peculiarity - a density increase at the rear edge of the ICME. In accordance with \citeA{rodriguez:2016}, we interpret this feature as follows: although the maximum plasma speed inside the ICME is higher than that of the trailing SW, which was one of our selection criteria, the velocity profile inside the MO exhibits a gradual decrease. Such a feature is commonly attributed to the ICME expansion \cite{Burlaga_Behannon1982}. Thus, the total velocity at the rear edge of the ejecta is lower than that of the trailing SW. The interaction of the ICME and the trailing SW leads to the compression of the plasma at the rear edge of the MO. A similar feature was observed in association with three more slow ICMEs.

In Figure~\ref{fig:analysis}) we show for each 12 hours interval (see Figure~\ref{fig:intervals}) the time-averaged properties of the SW and IMF before, during and after both ICMEs. The panels show a) magnetic field magnitude (nT), b) plasma number density (cm$^{-3}$), c) total SW velocity (kms$^{-1}$), d) SW dynamic pressure (nPa), e) temperature (K), f) IMF cone angle (degrees), g) thermal pressure (nPa) and h) magnetic pressure (nPa). On abscise of each panel in Figure~\ref{fig:analysis}A) we exhibit the number of the time intervals during which the SW and IMF properties were averaged. 
Note that at interval \#0 the derived peak values for the SW plasma and magnetic field occur in different regions of the ICME (shock-sheath and MO). The magnetic field magnitude peak is due to enhanced values inside the MO as well as sheath region, while the density peak is solely due to the sheath.

We first look at the July 2012 ICME. By comparing intervals directly preceding (interval \#$-$1) and following (interval \#$+$1) the event, we find that the most significant change ocurred for density and dynamic pressure. As detailed in Table~\ref{tab:2}, the IMF magnitude, plasma density, total velocity, dynamic pressure, and the cone angle just before the arrival of the ICME (interval \#$-$1) exhibit values of 4.54~nT, 4.8~cm$^{-3}$, 432~kms$^{-1}$, 1.49~nPa and 61$^\circ$, respectively. Just after the event (interval \#+1) these values were 4.59~nT, 1.1~cm$^{-3}$, 528~kms$^{-1}$, 0.51~nPa and 85$^\circ$. This means that the B-magnitude changed by 0.05~nT (1.0~\%), density by $-$3.7~cm$^{-3}$ ($-$77~\%), total velocity +98~kms$^{-1}$ ($+$22~\%), dynamic pressure by $-$0.98~nPa ($-$66~\%) and cone angle by $+$24$^\circ$.

We further inspect the SW parameters and IMF in the aftermath of the ICME and how the values change over time. In Table~\ref{tab:2} we show these values for intervals \#$-$1 and from \#$+$1 to \#$+$4 and the corresponding changes with respect to the interval \#$-$1. We can see that plasma density and dynamic pressure remained below the \#$-$1 values during 48 hours following the ICME. The velocity and cone angle exhibited increased values during the same time interval. The B-magnitude exhibits a slight increase during the interval \#+1 and a decrease for intervals \#+2--\#+4. 
With the exception of the cone angle, the behavior of the rest of the parameters agrees with the hypothesis that the ICME preconditioned the SW and IMF. If there was a successive ICME, even several days later, it would have been propagating in IP conditions of lower drag compared to that ICME studied.

In the case of the September 2010 event (Figure~\ref{fig:analysis}B) and Table~\ref{tab:2}) the magnetic field magnitude and SW density values diminished during the intervals \#+1--\#+4 compared to the values during the interval \#$-$1, while the SW velocity exhibited increased values during the same time period. The cone angle exhibited an initial increase (interval \#+1) followed by a decrease (intervals \#+2--\#+4).

\begin{center}
\begin{table}
\scalebox{0.9}{
\begin{tabular}{c |c c c c c| c c c c| c c c c c}
Interval \# & -1 & 1 & 2 & 3 & 4 & 1 & 2 & 3 & 4 & 1 & 2 & 3 & 4\\ 
\hline
\multicolumn{14}{c}{2--9 July 2012}\\
\hline
& \multicolumn{5}{c}{Value} & \multicolumn{4}{c}{Change} & \multicolumn{4}{c}{Change in \%} \\
 B (nT) & 4.54 & 4.59 & 3.33 & 4.11 & 4.33 & 0.05 & -1.21 & -0.43 & -0.21 & 1.0 & -27 & -9 & -5\\
N (cm$^{-3}$) & 4.80 & 1.1 & 1.28 & 1.94 & 4.19 & -3.7 & -3.52 & -2.86  & 0.61 & -77 & -73 & -60 & -13\\
V (kms$^{-1}$) & 432 & 528 & 511 & 482 & 448 & 98 & 79 &  50 & 16 & 22 & 18 & 12 & 4\\
T (10$^4$ K) & 8.10 & 8.23 & 6.9 & 6.7 & 4.8 & 1.29 & -1.18 & -1.39 & -3.2 & 2 & -15 & -17 & -40 & \\
P$_{dyn}$ (nPa) & 1.49 & 0.51 & 0.56 & 0.75 & 1.4 & -0.98 & -0.94 & -0.74 & -0.09 & -66 & -63 & -50 & -6\\
P$_{th}$ (10$^{-3}$ nPa) & 5.36 & 1.25 & 1.22 & 1.80 & 2.79 & -4.12 & -4.15 & -3.57 & -2.58& -78 & -77 & -66 & -48\\ 
P$_B$ (10$^{-2}$ nPa) & 1.64 & 1.67 & 0.88 & 1.35 & 1.49 & 0.034 & -0.76 & -0.29 & -0.15 & 21 & -46 & -18 & -9\\
IMF cone angle (deg) & 61 & 85 & 75 & 86 & 88 & 24 & 14 & 25 & 27 & & & & \\
\hline
\multicolumn{14}{c}{15--21 September 2010}\\
\hline
& \multicolumn{5}{c}{Value} & \multicolumn{4}{c}{Change} & \multicolumn{4}{c}{Change in \%} \\
 B (nT) & 5.24 & 4.69 & 4.04 & 3.50 & 3.25 & -0.55 & -1.20 & -1.74 & -1.99 & -11 & -22 & -33 & -38 \\
N (cm$^{-3}$) & 3.15 & 3.14 & 1.56 & 1.55 & 2.06 & -0.01 & -1.59 & -1.60 & -1.09 & -0.4 & -51 & -51 & -35 \\
V (kms$^{-1}$) & 336 & 390 & 359 & 340 & 341 & 55 & 24 & 5 & 5 & 16 & 7 & 1 & 2\\
T (10$^4$ K) & 8.33 & 10.01 & 4.856 & 4.06 & 4.61 & 1.67 & -3.48 & -4.27 & -3.7 & 20 & -42 & -51 & -45\\
P$_{dyn}$ (nPa) & 0.60 & 0.80 & 0.34 & 0.30 & 0.40 & 0.21 & -0.26 & -0.29 & -0.20 & 35 & -43 & -49 & -33\\
P$_{th}$ (10$^{-3}$ nPa) & 3.63 & 4.34 & 1.04 & 0.87 & 1.31 & 0.71 & -2.58 & -2.76 & -2.31 & 20 & -71 & -76 & -64 \\ 
P$_B$ (10$^{-2}$ nPa) & 2.18 & 1.75 & 1.30 & 0.98 & 0.84 & -0.43 & -0.88 & -1.21 &  -1.34 & -20 & -40 & -55 & -61\\
IMF cone angle (deg) & 61 & 77 & 37 & 51 & 56 & 16 & -23 & -10 & -5 & & & & \\
\hline
\end{tabular}
}
\caption{The first columns show time averaged values of B-magnitude, SW density, velocity, temperature, dynamic pressure, thermal pressure, IMF magnetic pressure and cone angle during the last time interval before the arrival of the 4--5 July 2012 and 17--19 September 2010 ICMEs (interval \#$-$1) and during the four intervals following the events. The next four columns show the change in the quantities with respect to the interval \#$-$1 and the last four columns show these changes in \%.}
\label{tab:2}
\end{table}
\end{center}

\subsection{Statistical analysis}
\begin{figure}
\centering
\includegraphics[width=1.2\textwidth]{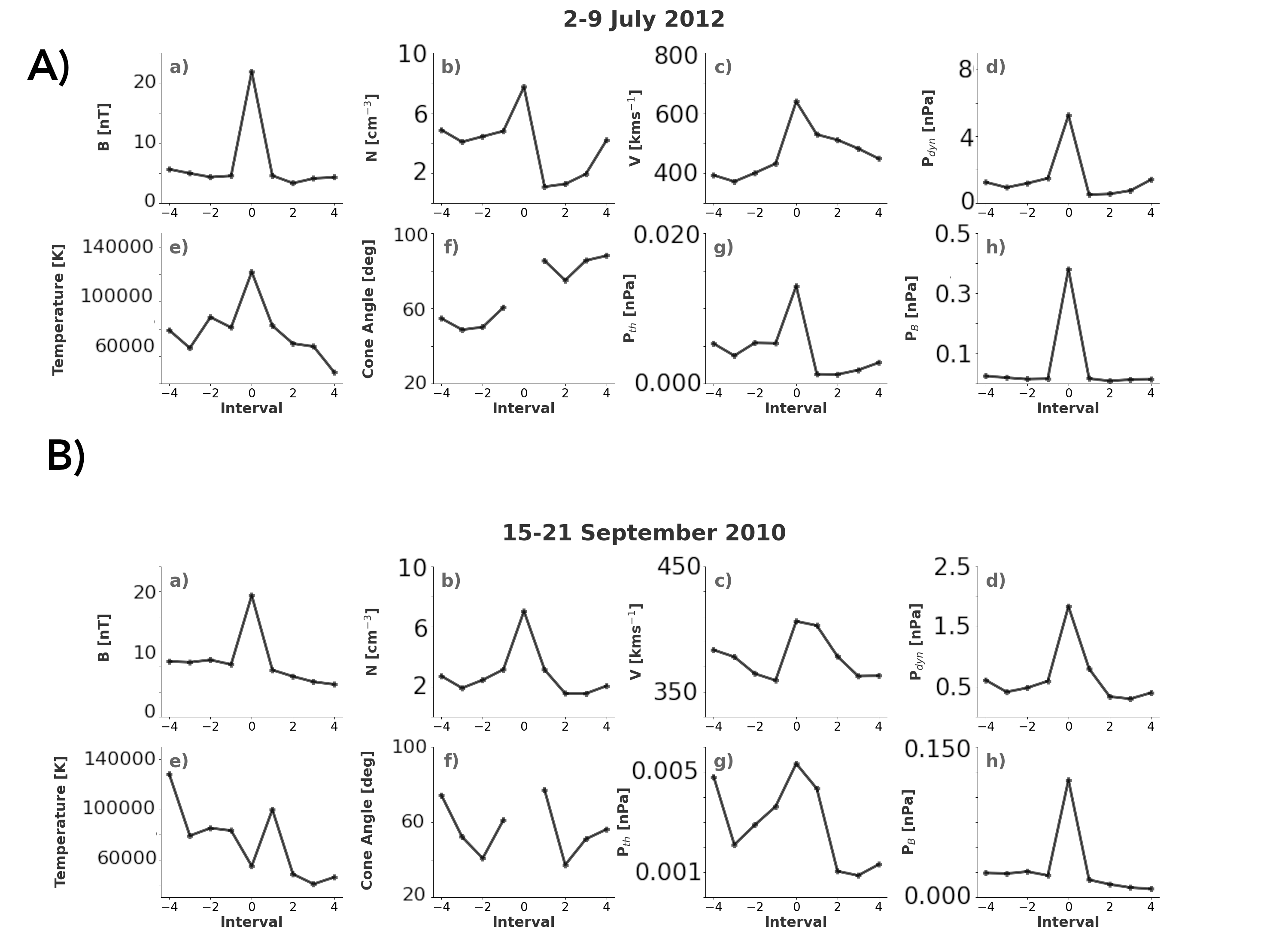}
\caption{Time averages of the SW and IMF properties during intervals before, during and after the 4--5 July 2012 and 17--19 September 2010 ICMEs. The panels show averages of a) magnetic field magnitude (in units of nanoTesla, nT), b) plasma number density (cm$^{-3}$), c) plasma total velocity (kms$^{-1}$), d)
dynamic pressure (nPa), e) temperature (K), f) cone angle (degrees), f) thermal pressure (nPa) and h) magnetic pressure (nPa). The intervals marked as 0 show averages during the entire period of the ICME. Points with negative sign correspond to successive 12 hour averages prior to the ICME arrival. Those with positive sign refer to 12 hour intervals after the ICME passage.}
\label{fig:analysis}
\end{figure}

\begin{figure}
\centering
\includegraphics[angle=0, width=1.2\textwidth]{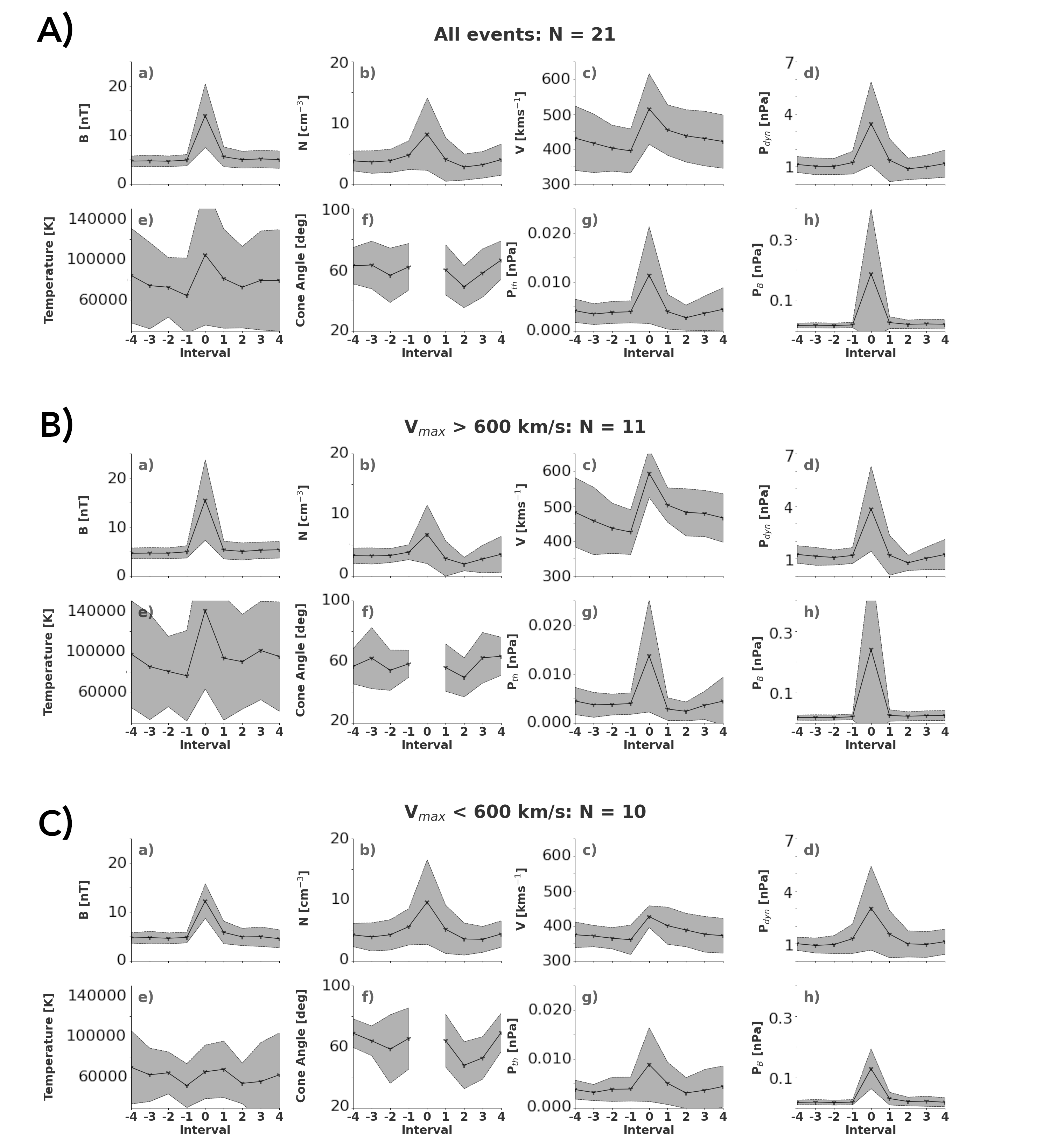}
\caption{This Figure has the same format as the Figure~\ref{fig:analysis}, only now  the average properties are shown for all (A), fast (B) and slow (C) events. The shaded areas represent the standard deviations of each averaged quantity.}
\label{fig:tendencies}
\end{figure}

In Figure~\ref{fig:tendencies}A) we plot the same profiles as in Figure~\ref{fig:analysis} but for all 21 events. The thick black trace represents the values averaged over all events in the sample. The shaded areas mark standard deviations ($\sigma$). The numerical values of the latter are provided in the Table~\ref{tab:appendixA} of the \ref{sec:appendixA} Section.

Since the number of events in our sample is relatively small, we conducted additional tests to assess whether $\sigma$ is an appropriate measure of variability for the values presented in Figure~\ref{fig:tendencies} and Table~\ref{tab:3}. Specifically, we applied the Shapiro-Wilk test \cite{shapiro:1965}, which is commonly used to evaluate whether a small sample (n $<$ 30) of univariate continuous data follows a normal distribution. The test yields a p-value: if this value is less than 0.05, the null hypothesis—that the sample is drawn from a normally distributed population—is rejected. As shown in Table A1, the p-values in most cases exceed the 0.05 threshold, indicating that the assumption of normality generally holds. However, there are a few exceptions, such as in the case of the temperature. To account for these cases, Table~\ref{tab:appendixA} also includes an alternative measure of statistical dispersion: the interquartile range (IQR), which is the difference between the third quartile (Q3) and the first quartile (Q1), corresponding to the 75th and 25th percentiles, respectively. Given that the majority of p-values are above 0.05, we proceed with using the mean and standard deviation as measures of central tendency and dispersion throughout the remainder of this study.

The results of this analysis are summarized in Table~\ref{tab:3} (top rows). These show that, on average, the region behind the ICMEs (interval \#+1--\#+4) tends to exhibit lower densities and smaller cone angles compared to the average values measured just before the arrival of the ICME (interval \#$-$1). The maximum differences are $-$41~\% and $-$13$^\circ$ for the density and cone angle, respectively. The trailing SW total velocity and magnetic field tend to exhibit somewhat higher values, by up to 15~\% and 14~\%, respectively. The behavior of the dynamic pressure is interesting since it seems to increase just behind the ICMEs, by 11~\% on average, but then it drops with respect to the immediate upstream value by between $-$4~\% and $-$29~\% on average. We can see that P$_{dyn}$ slowly recovers at later times. 

To see whether faster and thus more energetic ICMEs favor preconditioning of the IP medium, we divide our sample according to whether the maximum measured speed inside the events (during the interval \#0), V$_{\rm max}$, is lower or higher than 600~kms$^{-1}$. This is a somewhat arbitrary division, however, it gives us two subsamples covering a similar amount of events, namely 11 with V$_{\rm max}>$~600~kms$^{-1}$ and 10 events with V$_{\rm max}<$~600~kms$^{-1}$ .

Results for the faster ICMEs can be seen in Figure~\ref{fig:tendencies}B) and are summarized in Table~\ref{tab:3} (middle rows). We can see that the trends related to these events are similar to those of the whole sample, only that now the maximum variations of the B-magnitude, SW density, total velocity, P$_{dyn}$, and the IMF cone angle reach 7~\%, $-$50~\%, 18~\%, $-$36~\% and $-$9$^\circ$, respectively. The average P$_{dyn}$ initially remains almost the same, exhibiting a merely 1~\% rise, and then drops below the pre-ICME values.

We can compare these results with the values corresponding to slow ICMEs (Figure~\ref{fig:tendencies}C) and Table~\ref{tab:3}, bottom rows). The trends are similar as those for the entire sample. Changes of B-magnitude, SW density, total velocity, P$_{dyn}$ and the IMF cone angle reach 21~\%, $-$37~\%, 11~\%, $-$26~\% and $-$17$^\circ$, respectively. Again, the average P$_{dyn}$ first rises by 21~\% and then drops below the pre-ICME values.

\begin{center}
\begin{table}
\scalebox{0.9}{
\begin{tabular}{c |c c c c c| c c c c |c c c c }
\hline
\multicolumn{14}{c}{All events}\\
Interval \# & -1 & 1 & 2 & 3 & 4 & 1 & 2 & 3 & 4 & 1 & 2 & 3 & 4\\
\hline
& \multicolumn{5}{c}{Value} & \multicolumn{4}{c}{Change} & \multicolumn{4}{c}{Change in \%} \\
 B (nT) & 4.90& 5.58 & 5.00 & 5.14 & 5.00 & 0.68 & 0.10 & 0.25 & 0.11 & 14 & 2 & 5 & 2 \\
N (cm$^{-3}$) & 4.72 & 4.03 & 2.79 & 3.16 & 4.00 & -0.70 & -1.93 & -1.56 & -0.72 & -15 & -41 & -33 & -15 \\
V (kms$^{-1}$) & 395 & 455 & 438 & 430 & 422 & 60 & 43 & 35 & 27 & 15 & 11 & 9 & 7\\
T (10$^4$ K) & 6.48 & 8.14 & 7.31 & 7.97 & 7.97 & 1.66 & 8.24 & 1.49 & 1.49 & 26 & 13 & 23 & 23\\
P$_{dyn}$ (nPa) & 1.22 & 1.37 & 0.87 & 0.99 & 1.17 & 0.14 & -0.35 & -0.24 & -0.06 & 11 & -29 & -20 & -4\\
P$_{th}$ (10$^{-3}$ nPa) & 3.95 & 3.97 & 2.75 & 3.63 & 4.46 & 0.02 & -1.20 & -0.31 & 0.51 & -28 & -40 & -9 & -13\\ 
P$_B$ (10$^{-2}$ nPa) & 2.07 & 2.50 & 2.24 & 2.42 & 2.51 & 0.43 & 0.17 & 0.36 & 0.44 & 21 & 8 & 17 & 21\\
IMF cone angle (deg) & 62 & 60 & 49 & 58 & 66 & -2 & -13 & -4 & 4 & & & & \\
\hline
\multicolumn{14}{c}{}\\
\multicolumn{14}{c}{V$_{max}$ $>$ 600 kms$^{-1}$}\\
Interval \# & -1 & 1 & 2 & 3 & 4 & 1 & 2 & 3 & 4 & 1 & 2 & 3 & 4\\
\hline
& \multicolumn{5}{c}{Value} & \multicolumn{4}{c}{Change} & \multicolumn{4}{c}{Change in \%} \\
 B (nT) & 4.96 & 5.33 & 5.04 & 5.29 & 5.38 & 0.37 & 0.09 & 0.33 & 0.43 & 7 & 2 & 7 & 9 \\
N (cm$^{-3}$) & 3.91 & 2.98 & 1.97 & 2.80 & 3.57 & -1.04 & -1.94 & -1.11 & -0.34 & -27 & -50 & -28 & -9 \\
V (kms$^{-1}$) & 426 & 503 & 482 & 479 & 466 & 77 & 56 & 53 & 40 & 18 & 13 & 12 & 9\\
T (10$^4$ K) & 7.64 & 9.36 & 9.02 & 10.11 & 9.53 & 1.72 & 1.38 & 2.47 & 1.89 & 23 & 18 & 32 & 25\\
P$_{dyn}$ (nPa) & 1.18 & 1.19 & 0.76 & 1.02 & 1.24 & 0.01 & -0.42 & -0.16 & 0.05 & 1 & -36 & -14 & 5\\
P$_{th}$ (10$^{-3}$ nPa) & 3.99 & 2.87 & 2.40 & 3.63 & 4.49 & -1.12 & -1.59 & -0.37 & 0.50 & -28 & -40 & -9 & 13\\ 
 P$_B$ (10$^{-2}$ nPa) & 2.01 & 2.79 & 2.21 & 2.35 & 2.23 & 0.78 & 0.20 & 0.34 & 0.22 & 39 & 10 & 17 & 11 \\
IMF cone angle (deg) & 59 & 56 & 50 & 62 & 63 & -2 & -9 & 4 & 5 & & & & \\
\hline
\multicolumn{14}{c}{}\\
\multicolumn{14}{c}{V$_{max}$ $<$ 600 kms$^{-1}$}\\
Interval \# & -1 & 1 & 2 & 3 & 4 & 1 & 2 & 3 & 4 & 1 & 2 & 3 & 4\\
\hline
& \multicolumn{5}{c}{Value} & \multicolumn{4}{c}{Change} & \multicolumn{4}{c}{Change in \%} \\
 B (nT) & 4.93 & 5.86 & 4.94 & 4.87 & 4.58 & 1.03 & 0.11 & 0.16 & -0.25 & 21 & 2 & 3 & -5 \\
N (cm$^{-3}$) & 5.61 & 5.19 & 3.61 & 3.56 & 4.43 & -0.43 & -2.00 & -2.06 & -1.19 & -8 & -36 & -37 & -21 \\
V (kms$^{-1}$) & 361 & 401 & 389 & 377 & 373 & 40 & 28 & 16 & 12 & 11 & 8 & 4 & 3\\
T (10$^4$ K) & 5.21 & 6.80 & 5.42 & 5.61 & 6.26 & 1.59 & 0.22 & 0.40 & 1.05 & 31 & 4 & 8 & 20\\
P$_{dyn}$ (nPa) & 1.28 & 1.54 & 0.99 & 0.95 & 1.11 & 0.26 & -0.29 & -0.33 & -0.17 & 21 & -23 & -26 & -13\\
P$_{th}$ (10$^{-3}$ nPa) & 3.9 & 5.1 & 3.1 & 3.6 & 4.4 & 1.2 & -0.8 & -0.3 & 0.5 & 30 & -21 & -7 & 13 \\ 
P$_B$ (10$^{-2}$ nPa) & 1.94 & 3.11 & 2.17 & 2.26 & 1.91 & 1.17 & 0.22 &  0.32 & -0.03 & 60 & 11 & 16 & -2 \\
IMF cone angle (deg) & 65 & 64 & 48 & 53 & 69 & -1 & -17 & -12 & 4 & & & & \\
\hline
\end{tabular}
}
\caption{This table is in the same format as Table~\ref{tab:2} except that the exhibited values are averages over all 21 events (top rows), 11 fastest events with $V_{\rm max}>$~600~kms$^{-1}$ (middle rows) and 10 slowest events with top speeds $V_{\rm max}< 600$~kms$^{-1}$ (bottom rows).}\label{tab:3}
\end{table}
\end{center}

\section{Discussion}
\label{sec:discussion}
In this work we examine ICME catalogs from several space missions in order to search for isolated events and study their impact on the interplanetary medium. When a CME is ejected from the Sun, it propagates through the solar wind. If a previous ICME has already passed through a region, it can ``precondition'' that space, altering its density, pressure, and IMF structure. The goal is to determine and quantify the preconditioning effect of ICMEs on the SW plasma and IMF parameters. 

Preconditioning was invoked in the past in order to explain some of the most intense geomagnetic storms \cite<e.g.,>[]{knipp:2018, boteler:2019, hayakawa:2021} and the most energetic ICME ever detected in-situ \cite{liu:2014}.

Our definition of isolated events includes the requirement that the space mission did not detect any other SW structures, such as other ICMEs or CIRs during 48 hours prior to the arrival of the selected event and 48 during hours after their passage. Additionally, the ICMEs have to be faster than the trailing SW and they could not be complex events, e.g. two or more merged ICMEs or one ICME plus a CIR. Thus, we ended up with 21 events that fulfill these strict requirements.

Next, we examine the properties of the IMF and the SW prior, during and after the events. We do so by calculating during successive 12 hour time intervals the average B-magnitude, SW density, total velocity, temperature as well as dynamic pressure, magnetic pressure, and thermal pressure, and IMF cone angle. We then compare the post-ICME values with those just prior to the ICME arrival. We do this for all of the 21 events and then we separate them into two groups: those with maximum speed V$_{max} \geq$ 600~kms$^{-1}$ and those with V$_{max} < $600~kms$^{-1}$ to see whether there are any quantitative differences.

We find the following (see also the Table~\ref{tab:3}):
\begin{enumerate}
\item{} Post-ICME IMF magnitude and P$_B$ tend to exhibit higher values than prior ot the ICME arrival. In the case of faster events, this situation may last for at least 48 hours after the ICME passage. For slower ICMEs, this enhancement was observed only up to 36 hours following the events (see panel h) in Figure~\ref{fig:tendencies}).
\item{} The IMF cone angle, between the B-vector and the radial direction, just after the ICME (interval \#+1) tends to be similar to the one just before its arrival (interval \#$-$1). Afterwards the cone angle diminishes, and stays lower for the following 12 (fast events) or 24 (slow events) hours (panel f) in Figure~\ref{fig:tendencies}). 
\item{} The total velocity also increases after the ICME passage compared to the pre-ICME (interval \#$-$1) values. This enhancement was observed for up to 48 hours in both of our groups, but the fractional increase (in \%) was larger for slower events.
\item{} The post-ICME plasma density is lower than pre-ICME SW density. However, this decrease was largest for intervals \#+2 and \#+3, so between 12 and 24 and 24 and 36 hours after the end of the ICMEs as reported in the catalogs (panel c) in Figure~\ref{fig:tendencies}).
\item{} In the case of slow ICMEs the P$_{dyn}$ initially increases and decreases afterwards. In the case of fast events, the post-ICME dynamic pressure is initially similar to the pre-ICME P$_{dyn}$ and then it diminishes (panel d) in Figure~\ref{fig:tendencies}).
\item{} The post-ICME SW temperature exhibits higher values than the pre-ICME temperature for up to 48~hours. Its behavior is similar for fast and slow events(panel e) in Figure~\ref{fig:tendencies}).
\item{} Initial post-ICME values of the thermal pressure (P$_{th}$) are higher (lower) than the pre-ICME for slow (fast) events. During the period between 12 and 36 hrs after the ICME they are lower for both groups of events. During the interval \#+4 the SW P$_{th}$ associated with fast (slow) events exhibits larger (lower) values than the pre-ICME P$_{th}$ (panel g) in Figure~\ref{fig:tendencies}).

\end{enumerate}

Some of our findings agree with those from the already existing literature. \citeA{janvier:2019} studied magnetic field profiles of ICMEs and their surroundings at different heliospheric distances - at Mercury, Venus and Earth. The authors found that the post-ICME magnetic field intensities are larger than the pre-ICME ones and that this is especially true for slower ICMEs. Similar behavior is found in this study, as given in Table~\ref{tab:3}.

These authors also state that, independently of the heliocentric distance, the IMF magnitude in the wake of the events does not fully recover the pre-ICME properties and that the effects of ICMEs on the ambient SW last longer than the duration of the transient events. We can see in Table~\ref{tab:3} that the post-ICME B-magnitude exhibits enhanced values during 36 hours (slow events) and at least 48 hours (fast events) after the ICME passage. Our events lasted 1.74 days ($\sim$42~hours) on average, with the fast events lasting slightly less (1.72 days or 41.3 hours) compared to the slow events (1.76 days or 42.2 hours). This means that the average period of the enhanced post-ICME IMF magnitude is shorter (longer) than the duration of slow (fast) events. This probably indicates that, on average, it takes more time for the trailing SW to catch up with faster ICMEs and fill the wake behind them.

\citeA{temmer:2017} predicted that the SW velocity in the wake of the ICMEs should remain higher than expected for up to five days after the ICME start by between 9~\% and 24~\%. The average duration of events from their sample was $\sim$1.3 days, which means that the SW velocity remained enhanced for $\sim$3.7 days. From our Table~\ref{tab:3} we can see that the SW speed values in the wake of the ICMEs stayed above the pre-ICME values for at least 48 hours and that the average increase ranged between 7~\% and 18~\%. 

In the simulations of \citeA{wu:2022}, the region affected by the pre-events exhibited higher SW speed and diminished density. However it also exhibited diminished temperature while the IMF magnitude remained unchanged.

In order to explain the observed SW and IMF properties behind the ICMEs one should take into account their origin and conditions that they encounter in the interplanetary space. As the ejecta leave the Sun, the magnetic field and plasma that follow them originate from the same source active region. At later times, when the ICMEs reach larger heliocentric distances, the Sun rotates and the wake behind them is filled by the SW from regions adjacent to the active region of origin.

Soft X-ray and extreme ultraviolet (EUV) images show that the regions from which ICMEs originate exhibit so-called coronal dimmings, also referred to as “transient coronal holes” \cite{hansen:1974, rust:1983, hewish:1985, vanninathan:2018}. These are regions of diminished X-ray and EUV fluxes compared to the rest of the source active region. Relevant to this study are the secondary dimmings which are less intense, more diffuse and widespread than the primary ones \cite{mandrini:2007}. 
In general, dimmings are interpreted to be associated with open magnetic fields from the source region along which the hot coronal plasma escapes more freely into the IP space, leaving behind low-density areas. Plasma and B-field from the CME source region could be the reason for observed lower SW densities, higher velocities, temperatures and magnetic field magnitudes immediately behind ICMEs. In future studies, it would be valuable to relate the depletion in density measured in situ to the occurrence and evolution of dimming regions observed on the Sun \cite{dissauer:2019}. Furthermore, there are unique opportunities to directly compare the density depletion in a CME's wake, as recently observed from Parker Solar Probe's heliospheric white-light images \cite{stenborg:2023}.

The diminished SW densities further behind the ICMEs can be attributed to the ``snowplow'' effect, meaning that the ICMEs simply sweep out the solar wind, leaving more empty region of space behind them \cite<e.g., >[]{manchester:2004}. The fast SW streams that follow, propagate into an ``empty'' space and are able to maintain higher velocity at large heliocentric distances. The increased post-ICME SW temperature can be explained in the context of a well documented observational fact that it positively correlates with the SW speed \cite{burlaga:1973, lopez:1986, shi:2023}. Hence higher post-ICME velocities imply higher temperatures.

Higher post-ICME speeds may also account for more radial IMF: in the frozen-in  approximation, where the IMF is simply carried antisunward by the SW, the cone angle at a fixed heliocentric distance is inversely proportional to the SW speed. Hence, higher post-ICME speeds mean smaller cone angles (more radial Parker spiral).

The behavior of complex quantities, such as P$_{dyn}$ and P$_{th}$ is influenced by the behavior of two or more simple SW properties. For example, P$_{dyn}$ is a product of density and the square of the SW speed. Behind the fast ICMEs, the velocity increases and the density decrease initially compensate and P$_{dyn}$ values there are similar to the pre-ICME ones. In the case of slow ICMEs, the velocity increase prevails, so the P$_{dyn}$ increases. During the following 24 hours, the density decrease prevails and the P$_{dyn}$ decreases.

Similarly, P$_{th}$ depends on the SW density and temperature. In the case of fast (slow) ICMEs the density (temperature) decrease (increase) prevails, so during the interval \#+1 the P$_{th}$ decreases (increases) compared to the \#-1 values. During the following 24 hours the P$_{th}$ exhibits values that are lower than the pre-ICME values and then they increase during the interval \#+4.

Such post-ICME SW properties will influence propagation of any subsequent ejecta that might follow the initial events. Faster SW speeds and lower densities imply smaller drag force, as can be seen from equation~\ref{eq:drag}. Additionally, less dense SW increases the initial ratios between the ICME and the SW densities, $\rho_{\rm i}/\rho_{\rm e}$, which, as shown by \citeA{cargill:2004}, means lower deceleration rates.

Finally, we point to another feature visible in panels (c) and (e) of Figure~\ref{fig:tendencies}A): a gradual decrease of the SW velocity and temperature prior to ICMEs. This trend is particularly pronounced in the case of fast events (Figure~\ref{fig:tendencies}B), while it is much more subtle for slow ICMEs (Figure~\ref{fig:tendencies}C).

Although the origin of this tendency lies beyond the scope of the present study, it is worth noting that a gradual decrease in SW velocity appears in \citeA{masias:2016} (see their Figure 4). Interestingly, their results pertain to slow events, which contrasts with our observations. Further investigation based on a larger dataset will be needed to draw more definitive conclusions on this issue.

\section{Conclusions}
\label{sec:conclusions}
By comparing the properties of the SW and IMF prior and after the 21 isolated ICMEs we conclude the following:
\begin{enumerate}
\item{} On average, post-ICME IMF magnitude tends to exhibit higher values than the pre-ICME IMF.
\item{} The IMF cone angle during the first 12 hours after the ICME tends to be similar to the one observed during 12 hours prior to the ICME arrival. Afterwards the cone angle diminishes, and stays lower for the following 12 to 24 hours. 
\item{} The total velocity increases after the ICME passage compared to the pre-ICME values. This enhancement is observed for up to 48 hours following the ICMEs.
\item{} The post-ICME plasma density is lower than pre-ICME SW density. However, the maximum decrease does not occur immediately after the ICME passage, rather between 12 and 36 hours afterwards. 
\item{} During the first 12 hours after the ICME, the P$_{dyn}$ increases compared to the pre-ICME P$_{dyn}$. After that it drops significantly.
\item{} The ICMEs tend to modify the properties of the trailing IP medium in such a way, that they provide the IP conditions in which any following ICMEs propagating in a similar direction may do so more freely and experience less deceleration. Simulations performed by \citeA{temmer:2017} show that in the absence of any subsequent events, such favorable conditions may last for more than 5 days after the ICME observation.
\end{enumerate}

It is important to study the preconditioning of the IP medium by ICMEs for several reasons. For one, when ICMEs retain higher propagation speeds, they can cause more severe space weather events \cite<e.g., >{knipp:2018, boteler:2019, hayakawa:2021}. Also, such fast ejecta can interact with the preceding events, which can amplify their combined geoeffectiveness \cite<see, for example>{scolini:2020, cid:2023}. Combined effects of two or more ICMEs are thought to be responsible for $\sim$10~\% of the so called superstorms with D$_{\rm st}\leq$250~nT, classified by \citeA{meng:2019} as Type~II storms that exhibit several dips in their Dst profiles and occur more often during the solar wind maxima or the declining phases of the solar cycle. Future work will include the study of other phenomena related to ICMEs that might be affected by the preconditioning, such as the time profiles of solar energetic particles and Forbush decreases.

\newpage
\appendix
\section{SW and IMF averages, standard deviations and p-values}
\label{sec:appendixA}
\begin{table}[!h]
\centering
\begin{tabular}{c c c c c c}
\hline\\
\multicolumn{6}{c}{All events}\\
Quantity / Interval \# & $-$1 & 1 & 2 & 3 & 4\\
\hline\\
B [nT]& 4.90 & 5.58 & 4.99 & 5.14 & 5.00\\
$\sigma_{\rm B}$ [nT]& 1.15 & 2.03 & 1.71 & 1.78 & 1.76\\
p-value & 0.30 & 0.67 & 0.32 & 0.25 & 0.26\\
IQR [nT] & 1.96 & 2.22 & 2.80 & 2.75 & 2.64\\
\\
N cm$^{-3}$ & 4.72 & 4.03 & 2.79 & 3.16 & 4.00\\
$\sigma_{\rm N}$ cm$^{-3}$ & 2.34 & 3.56 & 2.13 & 2.17 & 2.54\\
p-value & 0.04 & 1.0 & 1.0 & 0.03 & 1.0\\
IQR cm$^{-3}$ & 2.4 & & & 2.2 & \\
\\
V [kms$^{-1}$] & 395 & 455 & 438 & 430 & 422\\
$\sigma_{\rm V}$ [kms$^{-1}$] & 63 & 72 & 75 & 78 & 76\\
p-value & 0.08 & 0.93 & 0.46 & 0.77 & 0.66\\
IQR [kms$^{-1}$] & 88, 76 & 71 & 84 & 62\\
\\
P$_{dyn}$ [nPa] & 1.23 & 1.37 & 0.87 & 0.99 & 1.17\\
$\sigma_{\rm P_{dyn}}$  [nPa]& 0.65 & 1.22 & 0.61 & 0.67 & 0.77\\
p-value & 0.03 & 1.0 & 1.0 & 0.20 & 1.0\\
IQR [nPa] & 0.63 & & & 0.67 & \\
\\
Cone angle [$^\circ$]& 61.92 & 59.97 & 48.97 & 57.95 & 66.38\\
$\sigma_{\rm CA}$ [$^\circ$]& 15.26 & 16.40 & 13.72 & 15.77 & 12.53\\
p-value & 0.12 & 0.41 & 0.25 & 0.20 & 0.07\\
IQR & 16.34 & 23.96 & 21.92 & 22.86 & 16.31\\
\\
T [10$^{4}$K]& 64.8 & 81.4 & 73.1 & 79.7 & 79.7\\
$\sigma_{\rm T}$ [10$^{4}$K] & 36.7 & 48.6 & 39.9 & 48.5 & 49.7\\
p-value & 0.03 & 0.01 & 0.003 & 0.21 & 0.03\\
IQR [10$^{4}$K] & 42 & 55 & 25 & 50 & 68\\
\\
P$_{\rm th}$ [10$^{-3}$ nPa]& 3.9 & 4.0 & 2.7 & 3.6 & 4.5\\
$\sigma_{\rm P_{\rm th}}$  [10$^{-3}$ nPa]& 2.3 & 3.6 & 2.6 & 3.5 & 4.4\\
p-value & 0.38 & 1.0 & 1.0 & 0.003 & 1.0\\
IQR [10$^{-3}$ nPa] & 3.7 & nan & nan & 3.4 & \\
\\
P$_{\rm B}$ [10$^{-3}$ nPa]& 20.1 & 27.9 & 22.1 & 23.5 & 22.3\\
$\sigma_{\rm P_{\rm B}}$  [10$^{-3}$ nPa]& 8.7 & 19.4 & 14.0 & 15.9 & 15.3\\
p-value & 0.34 & 0.045 & 0.13 & 0.03 & 0.03\\
IQR [10$^{-3}$ nPa] & 15.0 & 19.5 & 22.1 & 21.9 & 21.0\\
\hline
\end{tabular}
\caption{Average SW and IMF parameters together with the corresponding standard deviations, p-values from Shapiro-Wilk test and interquartile range (IQR) during intervals \#-1 (pre-ICME) and \#+1--\#+4 (post-ICME).}
\label{tab:appendixA}
\end{table}


\section*{Open Research Section}
We used the following resources:
\begin{itemize}
\item{} The Interplanetary Coronal Mass Ejections Multi-Catalog compiled by \citeA{larrodera:2024} available at\\ https://edatos.consorciomadrono.es/dataset.xhtml?persistentId=doi:10.21950/XGUIYX.
\item{} The Coordinated Data Analysis Web (CDAWeb, https://cdaweb.gsfc.nasa.gov/) portal for obtaining the data from STEREO, ACE and Helios missions.
\item{}  Some of the figures were produced with the ClWeb (https://clweb.irap.omp.eu/) on-line tool.
\end{itemize}

\acknowledgments
PK's work was funded by the DGAPA PAIIT IN100424 grant. MT acknowledges the financial support by the University of Graz. XBC's work was funded by CONAHCyT CBF2023-2024-852, and DGAPA PAPIIT IN106724 grants. We would like to thank Karin Dissauer for insightful discussions on statistical matters.


\begin{thebibliography}{}

\bibitem [\protect \citeauthoryear {%
Acu{\~{n}}a%
\ \protect \BOthers {.}}{%
Acu{\~{n}}a%
\ \protect \BOthers {.}}{%
{\protect \APACyear {2008}}%
}]{%
acuna:2008}
\APACinsertmetastar {%
acuna:2008}%
\begin{APACrefauthors}%
Acu{\~{n}}a, M\BPBI H.%
, Curtis, D.%
, Scheifele, J\BPBI L.%
, Russell, C\BPBI T.%
, Schroeder, P.%
, Szabo, A.%
\BCBL {}\ \BBA {} Luhmann, J\BPBI G.%
\end{APACrefauthors}%
\unskip\
\newblock
\APACrefYearMonthDay{2008}{Apr}{01}.
\newblock
{\BBOQ}\APACrefatitle {The STEREO/IMPACT Magnetic Field Experiment} {The
  stereo/impact magnetic field experiment}.{\BBCQ}
\newblock
\APACjournalVolNumPages{Space Science Reviews}{136}{1}{203--226}.
\newblock
\begin{APACrefURL} \url{https://doi.org/10.1007/s11214-007-9259-2}
  \end{APACrefURL}
\newblock
\begin{APACrefDOI} \doi{10.1007/s11214-007-9259-2} \end{APACrefDOI}
\PrintBackRefs{\CurrentBib}

\bibitem [\protect \citeauthoryear {%
{Bame}%
, {Asbridge}%
, {Feldman}%
, {Fenimore}%
\BCBL {}\ \BBA {} {Gosling}%
}{%
{Bame}%
\ \protect \BOthers {.}}{%
{\protect \APACyear {1979}}%
}]{%
bame:1979}
\APACinsertmetastar {%
bame:1979}%
\begin{APACrefauthors}%
{Bame}, S\BPBI J.%
, {Asbridge}, J\BPBI R.%
, {Feldman}, W\BPBI C.%
, {Fenimore}, E\BPBI E.%
\BCBL {}\ \BBA {} {Gosling}, J\BPBI T.%
\end{APACrefauthors}%
\unskip\
\newblock
\APACrefYearMonthDay{1979}{{\APACmonth{05}}}{}.
\newblock
{\BBOQ}\APACrefatitle {{Solar wind heavy ions from flare-heated coronal
  plasma.}} {{Solar wind heavy ions from flare-heated coronal plasma.}}{\BBCQ}
\newblock
\APACjournalVolNumPages{Solar Physics}{62}{1}{179-201}.
\newblock
\begin{APACrefDOI} \doi{10.1007/BF00150143} \end{APACrefDOI}
\PrintBackRefs{\CurrentBib}

\bibitem [\protect \citeauthoryear {%
Boteler%
}{%
Boteler%
}{%
{\protect \APACyear {2019}}%
}]{%
boteler:2019}
\APACinsertmetastar {%
boteler:2019}%
\begin{APACrefauthors}%
Boteler, D\BPBI H.%
\end{APACrefauthors}%
\unskip\
\newblock
\APACrefYearMonthDay{2019}{}{}.
\newblock
{\BBOQ}\APACrefatitle {A 21st Century View of the March 1989 Magnetic Storm} {A
  21st century view of the march 1989 magnetic storm}.{\BBCQ}
\newblock
\APACjournalVolNumPages{Space Weather}{17}{10}{1427-1441}.
\newblock
\begin{APACrefURL}
  \url{https://agupubs.onlinelibrary.wiley.com/doi/abs/10.1029/2019SW002278}
  \end{APACrefURL}
\newblock
\begin{APACrefDOI} \doi{https://doi.org/10.1029/2019SW002278} \end{APACrefDOI}
\PrintBackRefs{\CurrentBib}

\bibitem [\protect \citeauthoryear {%
L.~{Burlaga}%
, {Sittler}%
, {Mariani}%
\BCBL {}\ \BBA {} {Schwenn}%
}{%
L.~{Burlaga}%
\ \protect \BOthers {.}}{%
{\protect \APACyear {1981}}%
}]{%
burlaga:1981}
\APACinsertmetastar {%
burlaga:1981}%
\begin{APACrefauthors}%
{Burlaga}, L.%
, {Sittler}, E.%
, {Mariani}, F.%
\BCBL {}\ \BBA {} {Schwenn}, R.%
\end{APACrefauthors}%
\unskip\
\newblock
\APACrefYearMonthDay{1981}{{\APACmonth{08}}}{}.
\newblock
{\BBOQ}\APACrefatitle {{Magnetic loop behind an interplanetary shock: Voyager,
  Helios, and IMP 8 observations}} {{Magnetic loop behind an interplanetary
  shock: Voyager, Helios, and IMP 8 observations}}.{\BBCQ}
\newblock
\APACjournalVolNumPages{Journal of Geophysical Research: Space
  Physics}{86}{A8}{6673-6684}.
\newblock
\begin{APACrefDOI} \doi{10.1029/JA086iA08p06673} \end{APACrefDOI}
\PrintBackRefs{\CurrentBib}

\bibitem [\protect \citeauthoryear {%
L\BPBI F.~{Burlaga}%
\ \BBA {} {Behannon}%
}{%
L\BPBI F.~{Burlaga}%
\ \BBA {} {Behannon}%
}{%
{\protect \APACyear {1982}}%
}]{%
Burlaga_Behannon1982}
\APACinsertmetastar {%
Burlaga_Behannon1982}%
\begin{APACrefauthors}%
{Burlaga}, L\BPBI F.%
\BCBT {}\ \BBA {} {Behannon}, K\BPBI W.%
\end{APACrefauthors}%
\unskip\
\newblock
\APACrefYearMonthDay{1982}{{\APACmonth{11}}}{}.
\newblock
{\BBOQ}\APACrefatitle {{Magnetic Clouds - Voyager Observations Between 2AU and
  4AU}} {{Magnetic Clouds - Voyager Observations Between 2AU and 4AU}}.{\BBCQ}
\newblock
\APACjournalVolNumPages{\solphys}{81}{1}{181-192}.
\newblock
\begin{APACrefDOI} \doi{10.1007/BF00151989} \end{APACrefDOI}
\PrintBackRefs{\CurrentBib}

\bibitem [\protect \citeauthoryear {%
L\BPBI F.~{Burlaga}%
\ \BBA {} {Ogilvie}%
}{%
L\BPBI F.~{Burlaga}%
\ \BBA {} {Ogilvie}%
}{%
{\protect \APACyear {1973}}%
}]{%
burlaga:1973}
\APACinsertmetastar {%
burlaga:1973}%
\begin{APACrefauthors}%
{Burlaga}, L\BPBI F.%
\BCBT {}\ \BBA {} {Ogilvie}, K\BPBI W.%
\end{APACrefauthors}%
\unskip\
\newblock
\APACrefYearMonthDay{1973}{{\APACmonth{01}}}{}.
\newblock
{\BBOQ}\APACrefatitle {{Solar wind temperature and speed}} {{Solar wind
  temperature and speed}}.{\BBCQ}
\newblock
\APACjournalVolNumPages{Journal of Geophysical Research}{78}{13}{2028}.
\newblock
\begin{APACrefDOI} \doi{10.1029/JA078i013p02028} \end{APACrefDOI}
\PrintBackRefs{\CurrentBib}

\bibitem [\protect \citeauthoryear {%
L\BPBI F\BPBI E.~{Burlaga}%
}{%
L\BPBI F\BPBI E.~{Burlaga}%
}{%
{\protect \APACyear {1991}}%
}]{%
burlaga:1991}
\APACinsertmetastar {%
burlaga:1991}%
\begin{APACrefauthors}%
{Burlaga}, L\BPBI F\BPBI E.%
\end{APACrefauthors}%
\unskip\
\newblock
\APACrefYearMonthDay{1991}{}{}.
\newblock
{\BBOQ}\APACrefatitle {{Magnetic Clouds}} {{Magnetic Clouds}}.{\BBCQ}
\newblock
\BIn{} R.~{Schwenn}\ \BBA {} E.~{Marsch}\ (\BEDS), \APACrefbtitle {Physics of
  the Inner Heliosphere II} {Physics of the inner heliosphere ii}\ (\BVOL~21,
  \BPG~1-22).
\newblock
\begin{APACrefDOI} \doi{10.1007/978-3-642-75364-0_1} \end{APACrefDOI}
\PrintBackRefs{\CurrentBib}

\bibitem [\protect \citeauthoryear {%
{Cargill}%
}{%
{Cargill}%
}{%
{\protect \APACyear {2004}}%
}]{%
cargill:2004}
\APACinsertmetastar {%
cargill:2004}%
\begin{APACrefauthors}%
{Cargill}, P\BPBI J.%
\end{APACrefauthors}%
\unskip\
\newblock
\APACrefYearMonthDay{2004}{{\APACmonth{05}}}{}.
\newblock
{\BBOQ}\APACrefatitle {{On the Aerodynamic Drag Force Acting on Interplanetary
  Coronal Mass Ejections}} {{On the Aerodynamic Drag Force Acting on
  Interplanetary Coronal Mass Ejections}}.{\BBCQ}
\newblock
\APACjournalVolNumPages{Solar Physics}{221}{1}{135-149}.
\newblock
\begin{APACrefDOI} \doi{10.1023/B:SOLA.0000033366.10725.a2} \end{APACrefDOI}
\PrintBackRefs{\CurrentBib}

\bibitem [\protect \citeauthoryear {%
{Cargill}%
, {Chen}%
, {Spicer}%
\BCBL {}\ \BBA {} {Zalesak}%
}{%
{Cargill}%
\ \protect \BOthers {.}}{%
{\protect \APACyear {1996}}%
}]{%
cargill:1996}
\APACinsertmetastar {%
cargill:1996}%
\begin{APACrefauthors}%
{Cargill}, P\BPBI J.%
, {Chen}, J.%
, {Spicer}, D\BPBI S.%
\BCBL {}\ \BBA {} {Zalesak}, S\BPBI T.%
\end{APACrefauthors}%
\unskip\
\newblock
\APACrefYearMonthDay{1996}{{\APACmonth{03}}}{}.
\newblock
{\BBOQ}\APACrefatitle {{Magnetohydrodynamic simulations of the motion of
  magnetic flux tubes through a magnetized plasma}} {{Magnetohydrodynamic
  simulations of the motion of magnetic flux tubes through a magnetized
  plasma}}.{\BBCQ}
\newblock
\APACjournalVolNumPages{Journal of Geophysical Research}{101}{A3}{4855-4870}.
\newblock
\begin{APACrefDOI} \doi{10.1029/95JA03769} \end{APACrefDOI}
\PrintBackRefs{\CurrentBib}

\bibitem [\protect \citeauthoryear {%
{Cid}%
, {Saiz}%
, {Flores-Soriano}%
\BCBL {}\ \BBA {} {Knipp}%
}{%
{Cid}%
\ \protect \BOthers {.}}{%
{\protect \APACyear {2023}}%
}]{%
cid:2023}
\APACinsertmetastar {%
cid:2023}%
\begin{APACrefauthors}%
{Cid}, C.%
, {Saiz}, E.%
, {Flores-Soriano}, M.%
\BCBL {}\ \BBA {} {Knipp}, D\BPBI J.%
\end{APACrefauthors}%
\unskip\
\newblock
\APACrefYearMonthDay{2023}{{\APACmonth{12}}}{}.
\newblock
{\BBOQ}\APACrefatitle {{Interplanetary Signatures during the 1972 Early August
  Solar Storms}} {{Interplanetary Signatures during the 1972 Early August Solar
  Storms}}.{\BBCQ}
\newblock
\APACjournalVolNumPages{\apj}{958}{2}{159}.
\newblock
\begin{APACrefDOI} \doi{10.3847/1538-4357/acf9fd} \end{APACrefDOI}
\PrintBackRefs{\CurrentBib}

\bibitem [\protect \citeauthoryear {%
{Desai}%
\ \protect \BOthers {.}}{%
{Desai}%
\ \protect \BOthers {.}}{%
{\protect \APACyear {2020}}%
}]{%
desai:2020}
\APACinsertmetastar {%
desai:2020}%
\begin{APACrefauthors}%
{Desai}, R\BPBI T.%
, {Zhang}, H.%
, {Davies}, E\BPBI E.%
, {Stawarz}, J\BPBI E.%
, {Mico-Gomez}, J.%
\BCBL {}\ \BBA {} {Iv{\'a}{\~n}ez-Ballesteros}, P.%
\end{APACrefauthors}%
\unskip\
\newblock
\APACrefYearMonthDay{2020}{{\APACmonth{09}}}{}.
\newblock
{\BBOQ}\APACrefatitle {{Three-Dimensional Simulations of Solar Wind
  Preconditioning and the 23 July 2012 Interplanetary Coronal Mass Ejection}}
  {{Three-Dimensional Simulations of Solar Wind Preconditioning and the 23 July
  2012 Interplanetary Coronal Mass Ejection}}.{\BBCQ}
\newblock
\APACjournalVolNumPages{Solar Physics}{295}{9}{130}.
\newblock
\begin{APACrefDOI} \doi{10.1007/s11207-020-01700-5} \end{APACrefDOI}
\PrintBackRefs{\CurrentBib}

\bibitem [\protect \citeauthoryear {%
{Dissauer}%
, {Veronig}%
, {Temmer}%
\BCBL {}\ \BBA {} {Podladchikova}%
}{%
{Dissauer}%
\ \protect \BOthers {.}}{%
{\protect \APACyear {2019}}%
}]{%
dissauer:2019}
\APACinsertmetastar {%
dissauer:2019}%
\begin{APACrefauthors}%
{Dissauer}, K.%
, {Veronig}, A\BPBI M.%
, {Temmer}, M.%
\BCBL {}\ \BBA {} {Podladchikova}, T.%
\end{APACrefauthors}%
\unskip\
\newblock
\APACrefYearMonthDay{2019}{{\APACmonth{04}}}{}.
\newblock
{\BBOQ}\APACrefatitle {{Statistics of Coronal Dimmings Associated with Coronal
  Mass Ejections. II. Relationship between Coronal Dimmings and Their
  Associated CMEs}} {{Statistics of Coronal Dimmings Associated with Coronal
  Mass Ejections. II. Relationship between Coronal Dimmings and Their
  Associated CMEs}}.{\BBCQ}
\newblock
\APACjournalVolNumPages{\apj}{874}{2}{123}.
\newblock
\begin{APACrefDOI} \doi{10.3847/1538-4357/ab0962} \end{APACrefDOI}
\PrintBackRefs{\CurrentBib}

\bibitem [\protect \citeauthoryear {%
{Fisher}%
\ \BBA {} {Munro}%
}{%
{Fisher}%
\ \BBA {} {Munro}%
}{%
{\protect \APACyear {1984}}%
}]{%
fisher:1984}
\APACinsertmetastar {%
fisher:1984}%
\begin{APACrefauthors}%
{Fisher}, R\BPBI R.%
\BCBT {}\ \BBA {} {Munro}, R\BPBI H.%
\end{APACrefauthors}%
\unskip\
\newblock
\APACrefYearMonthDay{1984}{{\APACmonth{05}}}{}.
\newblock
{\BBOQ}\APACrefatitle {{Coronal transient geometry. I - The flare-associated
  event of 1981 March 25}} {{Coronal transient geometry. I - The
  flare-associated event of 1981 March 25}}.{\BBCQ}
\newblock
\APACjournalVolNumPages{\apj}{280}{}{428-439}.
\newblock
\begin{APACrefDOI} \doi{10.1086/162009} \end{APACrefDOI}
\PrintBackRefs{\CurrentBib}

\bibitem [\protect \citeauthoryear {%
{Gopalswamy}%
\ \protect \BOthers {.}}{%
{Gopalswamy}%
\ \protect \BOthers {.}}{%
{\protect \APACyear {2000}}%
}]{%
gopalswamy:2000}
\APACinsertmetastar {%
gopalswamy:2000}%
\begin{APACrefauthors}%
{Gopalswamy}, N.%
, {Lara}, A.%
, {Lepping}, R\BPBI P.%
, {Kaiser}, M\BPBI L.%
, {Berdichevsky}, D.%
\BCBL {}\ \BBA {} {St. Cyr}, O\BPBI C.%
\end{APACrefauthors}%
\unskip\
\newblock
\APACrefYearMonthDay{2000}{{\APACmonth{01}}}{}.
\newblock
{\BBOQ}\APACrefatitle {{Interplanetary acceleration of coronal mass ejections}}
  {{Interplanetary acceleration of coronal mass ejections}}.{\BBCQ}
\newblock
\APACjournalVolNumPages{Geophysical Research Letters}{27}{2}{145-148}.
\newblock
\begin{APACrefDOI} \doi{10.1029/1999GL003639} \end{APACrefDOI}
\PrintBackRefs{\CurrentBib}

\bibitem [\protect \citeauthoryear {%
Gosling%
\ \BBA {} Pizzo%
}{%
Gosling%
\ \BBA {} Pizzo%
}{%
{\protect \APACyear {1999}}%
}]{%
gosling:1999}
\APACinsertmetastar {%
gosling:1999}%
\begin{APACrefauthors}%
Gosling, J.%
\BCBT {}\ \BBA {} Pizzo, V.%
\end{APACrefauthors}%
\unskip\
\newblock
\APACrefYearMonthDay{1999}{}{}.
\newblock
{\BBOQ}\APACrefatitle {Formation and Evolution of Corotating Interaction
  Regions and their Three Dimensional Structure} {Formation and evolution of
  corotating interaction regions and their three dimensional structure}.{\BBCQ}
\newblock
\APACjournalVolNumPages{Space Science Reviews}{89}{1}{21--52}.
\newblock
\begin{APACrefURL} \url{http://dx.doi.org/10.1023/A:1005291711900}
  \end{APACrefURL}
\newblock
\begin{APACrefDOI} \doi{10.1023/A:1005291711900} \end{APACrefDOI}
\PrintBackRefs{\CurrentBib}

\bibitem [\protect \citeauthoryear {%
Hansen%
, Garcia%
, Hansen%
\BCBL {}\ \BBA {} Yasukawa%
}{%
Hansen%
\ \protect \BOthers {.}}{%
{\protect \APACyear {1974}}%
}]{%
hansen:1974}
\APACinsertmetastar {%
hansen:1974}%
\begin{APACrefauthors}%
Hansen, R\BPBI T.%
, Garcia, C\BPBI J.%
, Hansen, S\BPBI F.%
\BCBL {}\ \BBA {} Yasukawa, E.%
\end{APACrefauthors}%
\unskip\
\newblock
\APACrefYearMonthDay{1974}{aug}{}.
\newblock
{\BBOQ}\APACrefatitle {ABRUPT DEPLETIONS OF THE INNER CORONA} {Abrupt
  depletions of the inner corona}.{\BBCQ}
\newblock
\APACjournalVolNumPages{Publications of the Astronomical Society of the
  Pacific}{86}{512}{500}.
\newblock
\begin{APACrefURL} \url{https://dx.doi.org/10.1086/129638} \end{APACrefURL}
\newblock
\begin{APACrefDOI} \doi{10.1086/129638} \end{APACrefDOI}
\PrintBackRefs{\CurrentBib}

\bibitem [\protect \citeauthoryear {%
Hayakawa%
\ \protect \BOthers {.}}{%
Hayakawa%
\ \protect \BOthers {.}}{%
{\protect \APACyear {2021}}%
}]{%
hayakawa:2021}
\APACinsertmetastar {%
hayakawa:2021}%
\begin{APACrefauthors}%
Hayakawa, H.%
, Hattori, K.%
, Pevtsov, A\BPBI A.%
, Ebihara, Y.%
, Shea, M\BPBI A.%
, McCracken, K\BPBI G.%
\BDBL {}Knipp, D\BPBI J.%
\end{APACrefauthors}%
\unskip\
\newblock
\APACrefYearMonthDay{2021}{mar}{}.
\newblock
{\BBOQ}\APACrefatitle {The Intensity and Evolution of the Extreme Solar and
  Geomagnetic Storms in 1938 January} {The intensity and evolution of the
  extreme solar and geomagnetic storms in 1938 january}.{\BBCQ}
\newblock
\APACjournalVolNumPages{The Astrophysical Journal}{909}{2}{197}.
\newblock
\begin{APACrefURL} \url{https://dx.doi.org/10.3847/1538-4357/abc427}
  \end{APACrefURL}
\newblock
\begin{APACrefDOI} \doi{10.3847/1538-4357/abc427} \end{APACrefDOI}
\PrintBackRefs{\CurrentBib}

\bibitem [\protect \citeauthoryear {%
{Hewish}%
, {Tappin}%
\BCBL {}\ \BBA {} {Gapper}%
}{%
{Hewish}%
\ \protect \BOthers {.}}{%
{\protect \APACyear {1985}}%
}]{%
hewish:1985}
\APACinsertmetastar {%
hewish:1985}%
\begin{APACrefauthors}%
{Hewish}, A.%
, {Tappin}, S\BPBI J.%
\BCBL {}\ \BBA {} {Gapper}, G\BPBI R.%
\end{APACrefauthors}%
\unskip\
\newblock
\APACrefYearMonthDay{1985}{{\APACmonth{03}}}{}.
\newblock
{\BBOQ}\APACrefatitle {{Origin of strong interplanetary shocks}} {{Origin of
  strong interplanetary shocks}}.{\BBCQ}
\newblock
\APACjournalVolNumPages{Nature}{314}{6007}{137-140}.
\newblock
\begin{APACrefDOI} \doi{10.1038/314137a0} \end{APACrefDOI}
\PrintBackRefs{\CurrentBib}

\bibitem [\protect \citeauthoryear {%
{Janvier}%
\ \protect \BOthers {.}}{%
{Janvier}%
\ \protect \BOthers {.}}{%
{\protect \APACyear {2019}}%
}]{%
janvier:2019}
\APACinsertmetastar {%
janvier:2019}%
\begin{APACrefauthors}%
{Janvier}, M.%
, {Winslow}, R\BPBI M.%
, {Good}, S.%
, {Bonhomme}, E.%
, {D{\'e}moulin}, P.%
, {Dasso}, S.%
\BDBL {}{Boakes}, P\BPBI D.%
\end{APACrefauthors}%
\unskip\
\newblock
\APACrefYearMonthDay{2019}{{\APACmonth{02}}}{}.
\newblock
{\BBOQ}\APACrefatitle {{Generic Magnetic Field Intensity Profiles of
  Interplanetary Coronal Mass Ejections at Mercury, Venus, and Earth From
  Superposed Epoch Analyses}} {{Generic Magnetic Field Intensity Profiles of
  Interplanetary Coronal Mass Ejections at Mercury, Venus, and Earth From
  Superposed Epoch Analyses}}.{\BBCQ}
\newblock
\APACjournalVolNumPages{Journal of Geophysical Research (Space
  Physics)}{124}{2}{812-836}.
\newblock
\begin{APACrefDOI} \doi{10.1029/2018JA025949} \end{APACrefDOI}
\PrintBackRefs{\CurrentBib}

\bibitem [\protect \citeauthoryear {%
Kaiser%
\ \protect \BOthers {.}}{%
Kaiser%
\ \protect \BOthers {.}}{%
{\protect \APACyear {2008}}%
}]{%
kaiser:2008}
\APACinsertmetastar {%
kaiser:2008}%
\begin{APACrefauthors}%
Kaiser, M\BPBI L.%
, Kucera, T\BPBI A.%
, Davila, J\BPBI M.%
, St.~Cyr, O\BPBI C.%
, Guhathakurta, M.%
\BCBL {}\ \BBA {} Christian, E.%
\end{APACrefauthors}%
\unskip\
\newblock
\APACrefYearMonthDay{2008}{Apr}{01}.
\newblock
{\BBOQ}\APACrefatitle {The STEREO Mission: An Introduction} {The stereo
  mission: An introduction}.{\BBCQ}
\newblock
\APACjournalVolNumPages{Space Science Reviews}{136}{1}{5--16}.
\newblock
\begin{APACrefURL} \url{https://doi.org/10.1007/s11214-007-9277-0}
  \end{APACrefURL}
\newblock
\begin{APACrefDOI} \doi{10.1007/s11214-007-9277-0} \end{APACrefDOI}
\PrintBackRefs{\CurrentBib}

\bibitem [\protect \citeauthoryear {%
{Kajdi{\v{c}}}%
, {Alexandrova}%
, {Maksimovic}%
, {Lacombe}%
\BCBL {}\ \BBA {} {Fazakerley}%
}{%
{Kajdi{\v{c}}}%
\ \protect \BOthers {.}}{%
{\protect \APACyear {2016}}%
}]{%
kajdic:2016}
\APACinsertmetastar {%
kajdic:2016}%
\begin{APACrefauthors}%
{Kajdi{\v{c}}}, P.%
, {Alexandrova}, O.%
, {Maksimovic}, M.%
, {Lacombe}, C.%
\BCBL {}\ \BBA {} {Fazakerley}, A\BPBI N.%
\end{APACrefauthors}%
\unskip\
\newblock
\APACrefYearMonthDay{2016}{{\APACmonth{12}}}{}.
\newblock
{\BBOQ}\APACrefatitle {{Suprathermal Electron Strahl Widths in the Presence of
  Narrow-band Whistler Waves in the Solar Wind}} {{Suprathermal Electron Strahl
  Widths in the Presence of Narrow-band Whistler Waves in the Solar
  Wind}}.{\BBCQ}
\newblock
\APACjournalVolNumPages{\apj}{833}{2}{172}.
\newblock
\begin{APACrefDOI} \doi{10.3847/1538-4357/833/2/172} \end{APACrefDOI}
\PrintBackRefs{\CurrentBib}

\bibitem [\protect \citeauthoryear {%
{Kilpua}%
, {Koskinen}%
\BCBL {}\ \BBA {} {Pulkkinen}%
}{%
{Kilpua}%
\ \protect \BOthers {.}}{%
{\protect \APACyear {2017}}%
}]{%
kilpua:2017a}
\APACinsertmetastar {%
kilpua:2017a}%
\begin{APACrefauthors}%
{Kilpua}, E.%
, {Koskinen}, H\BPBI E\BPBI J.%
\BCBL {}\ \BBA {} {Pulkkinen}, T\BPBI I.%
\end{APACrefauthors}%
\unskip\
\newblock
\APACrefYearMonthDay{2017}{{\APACmonth{12}}}{}.
\newblock
{\BBOQ}\APACrefatitle {{Coronal mass ejections and their sheath regions in
  interplanetary space}} {{Coronal mass ejections and their sheath regions in
  interplanetary space}}.{\BBCQ}
\newblock
\APACjournalVolNumPages{Living Reviews in Solar Physics}{14}{1}{5}.
\newblock
\begin{APACrefDOI} \doi{10.1007/s41116-017-0009-6} \end{APACrefDOI}
\PrintBackRefs{\CurrentBib}

\bibitem [\protect \citeauthoryear {%
Knipp%
, Fraser%
, Shea%
\BCBL {}\ \BBA {} Smart%
}{%
Knipp%
\ \protect \BOthers {.}}{%
{\protect \APACyear {2018}}%
}]{%
knipp:2018}
\APACinsertmetastar {%
knipp:2018}%
\begin{APACrefauthors}%
Knipp, D\BPBI J.%
, Fraser, B\BPBI J.%
, Shea, M\BPBI A.%
\BCBL {}\ \BBA {} Smart, D\BPBI F.%
\end{APACrefauthors}%
\unskip\
\newblock
\APACrefYearMonthDay{2018}{}{}.
\newblock
{\BBOQ}\APACrefatitle {On the Little-Known Consequences of the 4 August 1972
  Ultra-Fast Coronal Mass Ejecta: Facts, Commentary, and Call to Action} {On
  the little-known consequences of the 4 august 1972 ultra-fast coronal mass
  ejecta: Facts, commentary, and call to action}.{\BBCQ}
\newblock
\APACjournalVolNumPages{Space Weather}{16}{11}{1635-1643}.
\newblock
\begin{APACrefURL}
  \url{https://agupubs.onlinelibrary.wiley.com/doi/abs/10.1029/2018SW002024}
  \end{APACrefURL}
\newblock
\begin{APACrefDOI} \doi{https://doi.org/10.1029/2018SW002024} \end{APACrefDOI}
\PrintBackRefs{\CurrentBib}

\bibitem [\protect \citeauthoryear {%
{Larrodera}%
\ \BBA {} {Temmer}%
}{%
{Larrodera}%
\ \BBA {} {Temmer}%
}{%
{\protect \APACyear {2024}}%
}]{%
larrodera:2024}
\APACinsertmetastar {%
larrodera:2024}%
\begin{APACrefauthors}%
{Larrodera}, C.%
\BCBT {}\ \BBA {} {Temmer}, M.%
\end{APACrefauthors}%
\unskip\
\newblock
\APACrefYearMonthDay{2024}{{\APACmonth{05}}}{}.
\newblock
{\BBOQ}\APACrefatitle {{Evolution of coronal mass ejections with and without
  sheaths from the inner to the outer heliosphere: Statistical investigation
  for 1975 to 2022}} {{Evolution of coronal mass ejections with and without
  sheaths from the inner to the outer heliosphere: Statistical investigation
  for 1975 to 2022}}.{\BBCQ}
\newblock
\APACjournalVolNumPages{Astronomy \& Astrophysics}{685}{}{A89}.
\newblock
\begin{APACrefDOI} \doi{10.1051/0004-6361/202348641} \end{APACrefDOI}
\PrintBackRefs{\CurrentBib}

\bibitem [\protect \citeauthoryear {%
{Liu}%
\ \protect \BOthers {.}}{%
{Liu}%
\ \protect \BOthers {.}}{%
{\protect \APACyear {2014}}%
}]{%
liu:2014}
\APACinsertmetastar {%
liu:2014}%
\begin{APACrefauthors}%
{Liu}, Y\BPBI D.%
, {Luhmann}, J\BPBI G.%
, {Kajdi{\v{c}}}, P.%
, {Kilpua}, E\BPBI K\BPBI J.%
, {Lugaz}, N.%
, {Nitta}, N\BPBI V.%
\BDBL {}{Galvin}, A\BPBI B.%
\end{APACrefauthors}%
\unskip\
\newblock
\APACrefYearMonthDay{2014}{{\APACmonth{03}}}{}.
\newblock
{\BBOQ}\APACrefatitle {{Observations of an extreme storm in interplanetary
  space caused by successive coronal mass ejections}} {{Observations of an
  extreme storm in interplanetary space caused by successive coronal mass
  ejections}}.{\BBCQ}
\newblock
\APACjournalVolNumPages{Nature Communications}{5}{}{3481}.
\newblock
\begin{APACrefDOI} \doi{10.1038/ncomms4481} \end{APACrefDOI}
\PrintBackRefs{\CurrentBib}

\bibitem [\protect \citeauthoryear {%
Liu%
, Zhao%
, Hu%
, Vourlidas%
\BCBL {}\ \BBA {} Zhu%
}{%
Liu%
\ \protect \BOthers {.}}{%
{\protect \APACyear {2019}}%
}]{%
liu:2019}
\APACinsertmetastar {%
liu:2019}%
\begin{APACrefauthors}%
Liu, Y\BPBI D.%
, Zhao, X.%
, Hu, H.%
, Vourlidas, A.%
\BCBL {}\ \BBA {} Zhu, B.%
\end{APACrefauthors}%
\unskip\
\newblock
\APACrefYearMonthDay{2019}{mar}{}.
\newblock
{\BBOQ}\APACrefatitle {A Comparative Study of 2017 July and 2012 July Complex
  Eruptions: Are Solar Superstorms “Perfect Storms” in Nature?} {A
  comparative study of 2017 july and 2012 july complex eruptions: Are solar
  superstorms “perfect storms” in nature?}{\BBCQ}
\newblock
\APACjournalVolNumPages{The Astrophysical Journal Supplement
  Series}{241}{2}{15}.
\newblock
\begin{APACrefURL} \url{https://dx.doi.org/10.3847/1538-4365/ab0649}
  \end{APACrefURL}
\newblock
\begin{APACrefDOI} \doi{10.3847/1538-4365/ab0649} \end{APACrefDOI}
\PrintBackRefs{\CurrentBib}

\bibitem [\protect \citeauthoryear {%
{Lopez}%
\ \BBA {} {Freeman}%
}{%
{Lopez}%
\ \BBA {} {Freeman}%
}{%
{\protect \APACyear {1986}}%
}]{%
lopez:1986}
\APACinsertmetastar {%
lopez:1986}%
\begin{APACrefauthors}%
{Lopez}, R\BPBI E.%
\BCBT {}\ \BBA {} {Freeman}, J\BPBI W.%
\end{APACrefauthors}%
\unskip\
\newblock
\APACrefYearMonthDay{1986}{{\APACmonth{02}}}{}.
\newblock
{\BBOQ}\APACrefatitle {{Solar wind proton temperature-velocity relationship}}
  {{Solar wind proton temperature-velocity relationship}}.{\BBCQ}
\newblock
\APACjournalVolNumPages{Journal of Geophysical Research}{91}{A2}{1701-1705}.
\newblock
\begin{APACrefDOI} \doi{10.1029/JA091iA02p01701} \end{APACrefDOI}
\PrintBackRefs{\CurrentBib}

\bibitem [\protect \citeauthoryear {%
Luhmann%
\ \protect \BOthers {.}}{%
Luhmann%
\ \protect \BOthers {.}}{%
{\protect \APACyear {2008}}%
}]{%
luhmann:2008}
\APACinsertmetastar {%
luhmann:2008}%
\begin{APACrefauthors}%
Luhmann, J\BPBI G.%
, Curtis, D\BPBI W.%
, Schroeder, P.%
, McCauley, J.%
, Lin, R\BPBI P.%
, Larson, D\BPBI E.%
\BDBL {}Gosling, J\BPBI T.%
\end{APACrefauthors}%
\unskip\
\newblock
\APACrefYearMonthDay{2008}{Apr}{01}.
\newblock
{\BBOQ}\APACrefatitle {STEREO IMPACT Investigation Goals, Measurements, and
  Data Products Overview} {Stereo impact investigation goals, measurements, and
  data products overview}.{\BBCQ}
\newblock
\APACjournalVolNumPages{Space Science Reviews}{136}{1}{117--184}.
\newblock
\begin{APACrefURL} \url{https://doi.org/10.1007/s11214-007-9170-x}
  \end{APACrefURL}
\newblock
\begin{APACrefDOI} \doi{10.1007/s11214-007-9170-x} \end{APACrefDOI}
\PrintBackRefs{\CurrentBib}

\bibitem [\protect \citeauthoryear {%
Manchester~IV%
\ \protect \BOthers {.}}{%
Manchester~IV%
\ \protect \BOthers {.}}{%
{\protect \APACyear {2004}}%
}]{%
manchester:2004}
\APACinsertmetastar {%
manchester:2004}%
\begin{APACrefauthors}%
Manchester~IV, W\BPBI B.%
, Gombosi, T\BPBI I.%
, Roussev, I.%
, De~Zeeuw, D\BPBI L.%
, Sokolov, I\BPBI V.%
, Powell, K\BPBI G.%
\BDBL {}Opher, M.%
\end{APACrefauthors}%
\unskip\
\newblock
\APACrefYearMonthDay{2004}{}{}.
\newblock
{\BBOQ}\APACrefatitle {Three-dimensional MHD simulation of a flux rope driven
  CME} {Three-dimensional mhd simulation of a flux rope driven cme}.{\BBCQ}
\newblock
\APACjournalVolNumPages{Journal of Geophysical Research: Space
  Physics}{109}{A1}{}.
\newblock
\begin{APACrefURL}
  \url{https://agupubs.onlinelibrary.wiley.com/doi/abs/10.1029/2002JA009672}
  \end{APACrefURL}
\newblock
\begin{APACrefDOI} \doi{https://doi.org/10.1029/2002JA009672} \end{APACrefDOI}
\PrintBackRefs{\CurrentBib}

\bibitem [\protect \citeauthoryear {%
{Mandrini}%
\ \protect \BOthers {.}}{%
{Mandrini}%
\ \protect \BOthers {.}}{%
{\protect \APACyear {2007}}%
}]{%
mandrini:2007}
\APACinsertmetastar {%
mandrini:2007}%
\begin{APACrefauthors}%
{Mandrini}, C\BPBI H.%
, {Nakwacki}, M\BPBI S.%
, {Attrill}, G.%
, {van Driel-Gesztelyi}, L.%
, {D{\'e}moulin}, P.%
, {Dasso}, S.%
\BCBL {}\ \BBA {} {Elliott}, H.%
\end{APACrefauthors}%
\unskip\
\newblock
\APACrefYearMonthDay{2007}{{\APACmonth{08}}}{}.
\newblock
{\BBOQ}\APACrefatitle {{Are CME-Related Dimmings Always a Simple Signature of
  Interplanetary Magnetic Cloud Footpoints?}} {{Are CME-Related Dimmings Always
  a Simple Signature of Interplanetary Magnetic Cloud Footpoints?}}{\BBCQ}
\newblock
\APACjournalVolNumPages{Solar Physics}{244}{1-2}{25-43}.
\newblock
\begin{APACrefDOI} \doi{10.1007/s11207-007-9020-8} \end{APACrefDOI}
\PrintBackRefs{\CurrentBib}

\bibitem [\protect \citeauthoryear {%
{Marubashi}%
}{%
{Marubashi}%
}{%
{\protect \APACyear {1986}}%
}]{%
marubashi:1986}
\APACinsertmetastar {%
marubashi:1986}%
\begin{APACrefauthors}%
{Marubashi}, K.%
\end{APACrefauthors}%
\unskip\
\newblock
\APACrefYearMonthDay{1986}{{\APACmonth{01}}}{}.
\newblock
{\BBOQ}\APACrefatitle {{Structure of the interplanetary magnetic clouds and
  their solar origins}} {{Structure of the interplanetary magnetic clouds and
  their solar origins}}.{\BBCQ}
\newblock
\APACjournalVolNumPages{Advances in Space Research}{6}{6}{335-338}.
\newblock
\begin{APACrefDOI} \doi{10.1016/0273-1177(86)90172-9} \end{APACrefDOI}
\PrintBackRefs{\CurrentBib}

\bibitem [\protect \citeauthoryear {%
{Mas{\'\i}as-Meza}%
, {Dasso}%
, {D{\'e}moulin}%
, {Rodriguez}%
\BCBL {}\ \BBA {} {Janvier}%
}{%
{Mas{\'\i}as-Meza}%
\ \protect \BOthers {.}}{%
{\protect \APACyear {2016}}%
}]{%
masias:2016}
\APACinsertmetastar {%
masias:2016}%
\begin{APACrefauthors}%
{Mas{\'\i}as-Meza}, J\BPBI J.%
, {Dasso}, S.%
, {D{\'e}moulin}, P.%
, {Rodriguez}, L.%
\BCBL {}\ \BBA {} {Janvier}, M.%
\end{APACrefauthors}%
\unskip\
\newblock
\APACrefYearMonthDay{2016}{{\APACmonth{08}}}{}.
\newblock
{\BBOQ}\APACrefatitle {{Superposed epoch study of ICME sub-structures near
  Earth and their effects on Galactic cosmic rays}} {{Superposed epoch study of
  ICME sub-structures near Earth and their effects on Galactic cosmic
  rays}}.{\BBCQ}
\newblock
\APACjournalVolNumPages{\aap}{592}{}{A118}.
\newblock
\begin{APACrefDOI} \doi{10.1051/0004-6361/201628571} \end{APACrefDOI}
\PrintBackRefs{\CurrentBib}

\bibitem [\protect \citeauthoryear {%
McComas%
\ \protect \BOthers {.}}{%
McComas%
\ \protect \BOthers {.}}{%
{\protect \APACyear {1998}}%
}]{%
mccomas:1998}
\APACinsertmetastar {%
mccomas:1998}%
\begin{APACrefauthors}%
McComas, D.%
, Bame, S.%
, Barker, P.%
, Feldman, W.%
, Phillips, J.%
, Riley, P.%
\BCBL {}\ \BBA {} Griffee, J.%
\end{APACrefauthors}%
\unskip\
\newblock
\APACrefYearMonthDay{1998}{Jul}{01}.
\newblock
{\BBOQ}\APACrefatitle {Solar Wind Electron Proton Alpha Monitor (SWEPAM) for
  the Advanced Composition Explorer} {Solar wind electron proton alpha monitor
  (swepam) for the advanced composition explorer}.{\BBCQ}
\newblock
\APACjournalVolNumPages{Space Science Reviews}{86}{1}{563--612}.
\newblock
\begin{APACrefURL} \url{https://doi.org/10.1023/A:1005040232597}
  \end{APACrefURL}
\newblock
\begin{APACrefDOI} \doi{10.1023/A:1005040232597} \end{APACrefDOI}
\PrintBackRefs{\CurrentBib}

\bibitem [\protect \citeauthoryear {%
{McComas}%
, {Gosling}%
, {Bame}%
, {Smith}%
\BCBL {}\ \BBA {} {Cane}%
}{%
{McComas}%
\ \protect \BOthers {.}}{%
{\protect \APACyear {1989}}%
}]{%
mccomas:1989}
\APACinsertmetastar {%
mccomas:1989}%
\begin{APACrefauthors}%
{McComas}, D\BPBI J.%
, {Gosling}, J\BPBI T.%
, {Bame}, S\BPBI J.%
, {Smith}, E\BPBI J.%
\BCBL {}\ \BBA {} {Cane}, H\BPBI V.%
\end{APACrefauthors}%
\unskip\
\newblock
\APACrefYearMonthDay{1989}{{\APACmonth{02}}}{}.
\newblock
{\BBOQ}\APACrefatitle {{A test of magnetic field draping induced B$_{z}$
  perturbations ahead of fast coronal mass ejecta}} {{A test of magnetic field
  draping induced B$_{z}$ perturbations ahead of fast coronal mass
  ejecta}}.{\BBCQ}
\newblock
\APACjournalVolNumPages{Journal of Geophysical Research}{94}{A2}{1465-1471}.
\newblock
\begin{APACrefDOI} \doi{10.1029/JA094iA02p01465} \end{APACrefDOI}
\PrintBackRefs{\CurrentBib}

\bibitem [\protect \citeauthoryear {%
Meng%
, Tsurutani%
\BCBL {}\ \BBA {} Mannucci%
}{%
Meng%
\ \protect \BOthers {.}}{%
{\protect \APACyear {2019}}%
}]{%
meng:2019}
\APACinsertmetastar {%
meng:2019}%
\begin{APACrefauthors}%
Meng, X.%
, Tsurutani, B\BPBI T.%
\BCBL {}\ \BBA {} Mannucci, A\BPBI J.%
\end{APACrefauthors}%
\unskip\
\newblock
\APACrefYearMonthDay{2019}{}{}.
\newblock
{\BBOQ}\APACrefatitle {The Solar and Interplanetary Causes of Superstorms
  During the Space Age} {The solar and interplanetary causes of superstorms
  during the space age}.{\BBCQ}
\newblock
\APACjournalVolNumPages{Journal of Geophysical Research: Space
  Physics}{124}{6}{3926-3948}.
\newblock
\begin{APACrefURL}
  \url{https://agupubs.onlinelibrary.wiley.com/doi/abs/10.1029/2018JA026425}
  \end{APACrefURL}
\newblock
\begin{APACrefDOI} \doi{https://doi.org/10.1029/2018JA026425} \end{APACrefDOI}
\PrintBackRefs{\CurrentBib}

\bibitem [\protect \citeauthoryear {%
{Porsche}%
}{%
{Porsche}%
}{%
{\protect \APACyear {1981}}%
}]{%
porsche:1981}
\APACinsertmetastar {%
porsche:1981}%
\begin{APACrefauthors}%
{Porsche}, H.%
\end{APACrefauthors}%
\unskip\
\newblock
\APACrefYearMonthDay{1981}{{\APACmonth{11}}}{}.
\newblock
{\BBOQ}\APACrefatitle {{HELIOS mission: Mission objectives, mission
  verification, selected results}} {{HELIOS mission: Mission objectives,
  mission verification, selected results}}.{\BBCQ}
\newblock
\BIn{} W\BPBI R.~{Burke}\ (\BED), \APACrefbtitle {Solar System and its
  Exploration} {Solar system and its exploration}\ (\BVOL~164, \BPG~43-50).
\PrintBackRefs{\CurrentBib}

\bibitem [\protect \citeauthoryear {%
Reiss%
\ \protect \BOthers {.}}{%
Reiss%
\ \protect \BOthers {.}}{%
{\protect \APACyear {2016}}%
}]{%
reiss:2016}
\APACinsertmetastar {%
reiss:2016}%
\begin{APACrefauthors}%
Reiss, M\BPBI A.%
, Temmer, M.%
, Veronig, A\BPBI M.%
, Nikolic, L.%
, Vennerstrom, S.%
, Schöngassner, F.%
\BCBL {}\ \BBA {} Hofmeister, S\BPBI J.%
\end{APACrefauthors}%
\unskip\
\newblock
\APACrefYearMonthDay{2016}{}{}.
\newblock
{\BBOQ}\APACrefatitle {Verification of high-speed solar wind stream forecasts
  using operational solar wind models} {Verification of high-speed solar wind
  stream forecasts using operational solar wind models}.{\BBCQ}
\newblock
\APACjournalVolNumPages{Space Weather}{14}{7}{495-510}.
\newblock
\begin{APACrefURL}
  \url{https://agupubs.onlinelibrary.wiley.com/doi/abs/10.1002/2016SW001390}
  \end{APACrefURL}
\newblock
\begin{APACrefDOI} \doi{https://doi.org/10.1002/2016SW001390} \end{APACrefDOI}
\PrintBackRefs{\CurrentBib}

\bibitem [\protect \citeauthoryear {%
{Rodriguez}%
\ \protect \BOthers {.}}{%
{Rodriguez}%
\ \protect \BOthers {.}}{%
{\protect \APACyear {2016}}%
}]{%
rodriguez:2016}
\APACinsertmetastar {%
rodriguez:2016}%
\begin{APACrefauthors}%
{Rodriguez}, L.%
, {Mas{\'\i}as-Meza}, J\BPBI J.%
, {Dasso}, S.%
, {D{\'e}moulin}, P.%
, {Zhukov}, A\BPBI N.%
, {Gulisano}, A\BPBI M.%
\BDBL {}{Janvier}, M.%
\end{APACrefauthors}%
\unskip\
\newblock
\APACrefYearMonthDay{2016}{{\APACmonth{08}}}{}.
\newblock
{\BBOQ}\APACrefatitle {{Typical Profiles and Distributions of Plasma and
  Magnetic Field Parameters in Magnetic Clouds at 1 AU}} {{Typical Profiles and
  Distributions of Plasma and Magnetic Field Parameters in Magnetic Clouds at 1
  AU}}.{\BBCQ}
\newblock
\APACjournalVolNumPages{Solar Physics}{291}{7}{2145-2163}.
\newblock
\begin{APACrefDOI} \doi{10.1007/s11207-016-0955-5} \end{APACrefDOI}
\PrintBackRefs{\CurrentBib}

\bibitem [\protect \citeauthoryear {%
{Rust}%
}{%
{Rust}%
}{%
{\protect \APACyear {1983}}%
}]{%
rust:1983}
\APACinsertmetastar {%
rust:1983}%
\begin{APACrefauthors}%
{Rust}, D\BPBI M.%
\end{APACrefauthors}%
\unskip\
\newblock
\APACrefYearMonthDay{1983}{{\APACmonth{01}}}{}.
\newblock
{\BBOQ}\APACrefatitle {{Coronal Disturbances and Their Terrestrial Effects}}
  {{Coronal Disturbances and Their Terrestrial Effects}}.{\BBCQ}
\newblock
\APACjournalVolNumPages{Space Science Reviews}{34}{1}{21-36}.
\newblock
\begin{APACrefDOI} \doi{10.1007/BF00221193} \end{APACrefDOI}
\PrintBackRefs{\CurrentBib}

\bibitem [\protect \citeauthoryear {%
Schwenn%
}{%
Schwenn%
}{%
{\protect \APACyear {2006}}%
}]{%
schwenn:2006}
\APACinsertmetastar {%
schwenn:2006}%
\begin{APACrefauthors}%
Schwenn, R.%
\end{APACrefauthors}%
\unskip\
\newblock
\APACrefYearMonthDay{2006}{}{}.
\newblock
{\BBOQ}\APACrefatitle {Space Weather: The Solar Perspective} {Space weather:
  The solar perspective}.{\BBCQ}
\newblock
\APACjournalVolNumPages{Living Reviews in Solar Physics}{3}{1}{2}.
\newblock
\begin{APACrefURL} \url{http://dx.doi.org/10.12942/lrsp-2006-2}
  \end{APACrefURL}
\newblock
\begin{APACrefDOI} \doi{10.12942/lrsp-2006-2} \end{APACrefDOI}
\PrintBackRefs{\CurrentBib}

\bibitem [\protect \citeauthoryear {%
{Scolini}%
\ \protect \BOthers {.}}{%
{Scolini}%
\ \protect \BOthers {.}}{%
{\protect \APACyear {2020}}%
}]{%
scolini:2020}
\APACinsertmetastar {%
scolini:2020}%
\begin{APACrefauthors}%
{Scolini}, C.%
, {Chan{\'e}}, E.%
, {Temmer}, M.%
, {Kilpua}, E\BPBI K\BPBI J.%
, {Dissauer}, K.%
, {Veronig}, A\BPBI M.%
\BDBL {}{Poedts}, S.%
\end{APACrefauthors}%
\unskip\
\newblock
\APACrefYearMonthDay{2020}{{\APACmonth{02}}}{}.
\newblock
{\BBOQ}\APACrefatitle {{CME-CME Interactions as Sources of CME
  Geoeffectiveness: The Formation of the Complex Ejecta and Intense Geomagnetic
  Storm in 2017 Early September}} {{CME-CME Interactions as Sources of CME
  Geoeffectiveness: The Formation of the Complex Ejecta and Intense Geomagnetic
  Storm in 2017 Early September}}.{\BBCQ}
\newblock
\APACjournalVolNumPages{The Astrophysical Journal Supplement
  Series}{247}{1}{21}.
\newblock
\begin{APACrefDOI} \doi{10.3847/1538-4365/ab6216} \end{APACrefDOI}
\PrintBackRefs{\CurrentBib}

\bibitem [\protect \citeauthoryear {%
Shapiro%
\ \BBA {} Wilk%
}{%
Shapiro%
\ \BBA {} Wilk%
}{%
{\protect \APACyear {1965}}%
}]{%
shapiro:1965}
\APACinsertmetastar {%
shapiro:1965}%
\begin{APACrefauthors}%
Shapiro, S\BPBI S.%
\BCBT {}\ \BBA {} Wilk, M\BPBI B.%
\end{APACrefauthors}%
\unskip\
\newblock
\APACrefYearMonthDay{1965}{}{}.
\newblock
{\BBOQ}\APACrefatitle {An Analysis of Variance Test for Normality (Complete
  Samples)} {An analysis of variance test for normality (complete
  samples)}.{\BBCQ}
\newblock
\APACjournalVolNumPages{Biometrika}{52}{3/4}{591--611}.
\newblock
\begin{APACrefURL} [{2025-05-06}]\url{http://www.jstor.org/stable/2333709}
  \end{APACrefURL}
\PrintBackRefs{\CurrentBib}

\bibitem [\protect \citeauthoryear {%
{Shi}%
\ \protect \BOthers {.}}{%
{Shi}%
\ \protect \BOthers {.}}{%
{\protect \APACyear {2023}}%
}]{%
shi:2023}
\APACinsertmetastar {%
shi:2023}%
\begin{APACrefauthors}%
{Shi}, C.%
, {Velli}, M.%
, {Lionello}, R.%
, {Sioulas}, N.%
, {Huang}, Z.%
, {Halekas}, J\BPBI S.%
\BDBL {}{Bale}, S\BPBI D.%
\end{APACrefauthors}%
\unskip\
\newblock
\APACrefYearMonthDay{2023}{{\APACmonth{02}}}{}.
\newblock
{\BBOQ}\APACrefatitle {{Proton and Electron Temperatures in the Solar Wind and
  Their Correlations with the Solar Wind Speed}} {{Proton and Electron
  Temperatures in the Solar Wind and Their Correlations with the Solar Wind
  Speed}}.{\BBCQ}
\newblock
\APACjournalVolNumPages{The Astrophysical Journal}{944}{1}{82}.
\newblock
\begin{APACrefDOI} \doi{10.3847/1538-4357/acb341} \end{APACrefDOI}
\PrintBackRefs{\CurrentBib}

\bibitem [\protect \citeauthoryear {%
Smith%
\ \protect \BOthers {.}}{%
Smith%
\ \protect \BOthers {.}}{%
{\protect \APACyear {1998}}%
}]{%
smith:1998}
\APACinsertmetastar {%
smith:1998}%
\begin{APACrefauthors}%
Smith, C.%
, L'Heureux, J.%
, Ness, N.%
, Acu{\~{n}}a, M.%
, Burlaga, L.%
\BCBL {}\ \BBA {} Scheifele, J.%
\end{APACrefauthors}%
\unskip\
\newblock
\APACrefYearMonthDay{1998}{Jul}{01}.
\newblock
{\BBOQ}\APACrefatitle {The ACE Magnetic Fields Experiment} {The ace magnetic
  fields experiment}.{\BBCQ}
\newblock
\APACjournalVolNumPages{Space Science Reviews}{86}{1}{613--632}.
\newblock
\begin{APACrefURL} \url{https://doi.org/10.1023/A:1005092216668}
  \end{APACrefURL}
\newblock
\begin{APACrefDOI} \doi{10.1023/A:1005092216668} \end{APACrefDOI}
\PrintBackRefs{\CurrentBib}

\bibitem [\protect \citeauthoryear {%
{Stenborg}%
, {Paouris}%
, {Howard}%
, {Vourlidas}%
\BCBL {}\ \BBA {} {Hess}%
}{%
{Stenborg}%
\ \protect \BOthers {.}}{%
{\protect \APACyear {2023}}%
}]{%
stenborg:2023}
\APACinsertmetastar {%
stenborg:2023}%
\begin{APACrefauthors}%
{Stenborg}, G.%
, {Paouris}, E.%
, {Howard}, R\BPBI A.%
, {Vourlidas}, A.%
\BCBL {}\ \BBA {} {Hess}, P.%
\end{APACrefauthors}%
\unskip\
\newblock
\APACrefYearMonthDay{2023}{{\APACmonth{06}}}{}.
\newblock
{\BBOQ}\APACrefatitle {{Investigating Coronal Holes and CMEs as Sources of
  Brightness Depletion Detected in PSP/WISPR Images}} {{Investigating Coronal
  Holes and CMEs as Sources of Brightness Depletion Detected in PSP/WISPR
  Images}}.{\BBCQ}
\newblock
\APACjournalVolNumPages{\apj}{949}{2}{61}.
\newblock
\begin{APACrefDOI} \doi{10.3847/1538-4357/acd2cf} \end{APACrefDOI}
\PrintBackRefs{\CurrentBib}

\bibitem [\protect \citeauthoryear {%
Stone%
\ \protect \BOthers {.}}{%
Stone%
\ \protect \BOthers {.}}{%
{\protect \APACyear {1998}}%
}]{%
stone:1998}
\APACinsertmetastar {%
stone:1998}%
\begin{APACrefauthors}%
Stone, E.%
, Frandsen, A.%
, Mewaldt, R.%
, Christian, E.%
, Margolies, D.%
, Ormes, J.%
\BCBL {}\ \BBA {} Snow, F.%
\end{APACrefauthors}%
\unskip\
\newblock
\APACrefYearMonthDay{1998}{Jul}{01}.
\newblock
{\BBOQ}\APACrefatitle {The Advanced Composition Explorer} {The advanced
  composition explorer}.{\BBCQ}
\newblock
\APACjournalVolNumPages{Space Science Reviews}{86}{1}{1--22}.
\newblock
\begin{APACrefURL} \url{https://doi.org/10.1023/A:1005082526237}
  \end{APACrefURL}
\newblock
\begin{APACrefDOI} \doi{10.1023/A:1005082526237} \end{APACrefDOI}
\PrintBackRefs{\CurrentBib}

\bibitem [\protect \citeauthoryear {%
{Temmer}%
\ \BBA {} {Nitta}%
}{%
{Temmer}%
\ \BBA {} {Nitta}%
}{%
{\protect \APACyear {2015}}%
}]{%
temmer:2015}
\APACinsertmetastar {%
temmer:2015}%
\begin{APACrefauthors}%
{Temmer}, M.%
\BCBT {}\ \BBA {} {Nitta}, N\BPBI V.%
\end{APACrefauthors}%
\unskip\
\newblock
\APACrefYearMonthDay{2015}{{\APACmonth{03}}}{}.
\newblock
{\BBOQ}\APACrefatitle {{Interplanetary Propagation Behavior of the Fast Coronal
  Mass Ejection on 23 July 2012}} {{Interplanetary Propagation Behavior of the
  Fast Coronal Mass Ejection on 23 July 2012}}.{\BBCQ}
\newblock
\APACjournalVolNumPages{Solar Physics}{290}{3}{919-932}.
\newblock
\begin{APACrefDOI} \doi{10.1007/s11207-014-0642-3} \end{APACrefDOI}
\PrintBackRefs{\CurrentBib}

\bibitem [\protect \citeauthoryear {%
Temmer%
, Reiss%
, Nikolic%
, Hofmeister%
\BCBL {}\ \BBA {} Veronig%
}{%
Temmer%
\ \protect \BOthers {.}}{%
{\protect \APACyear {2017}}%
}]{%
temmer:2017}
\APACinsertmetastar {%
temmer:2017}%
\begin{APACrefauthors}%
Temmer, M.%
, Reiss, M\BPBI A.%
, Nikolic, L.%
, Hofmeister, S\BPBI J.%
\BCBL {}\ \BBA {} Veronig, A\BPBI M.%
\end{APACrefauthors}%
\unskip\
\newblock
\APACrefYearMonthDay{2017}{jan}{}.
\newblock
{\BBOQ}\APACrefatitle {Preconditioning of Interplanetary Space Due to Transient
  CME Disturbances} {Preconditioning of interplanetary space due to transient
  cme disturbances}.{\BBCQ}
\newblock
\APACjournalVolNumPages{The Astrophysical Journal}{835}{2}{141}.
\newblock
\begin{APACrefURL} \url{https://dx.doi.org/10.3847/1538-4357/835/2/141}
  \end{APACrefURL}
\newblock
\begin{APACrefDOI} \doi{10.3847/1538-4357/835/2/141} \end{APACrefDOI}
\PrintBackRefs{\CurrentBib}

\bibitem [\protect \citeauthoryear {%
{Tsurutani}%
, {Gonzalez}%
, {Tang}%
, {Akasofu}%
\BCBL {}\ \BBA {} {Smith}%
}{%
{Tsurutani}%
\ \protect \BOthers {.}}{%
{\protect \APACyear {1988}}%
}]{%
tsurutani:1988}
\APACinsertmetastar {%
tsurutani:1988}%
\begin{APACrefauthors}%
{Tsurutani}, B\BPBI T.%
, {Gonzalez}, W\BPBI D.%
, {Tang}, F.%
, {Akasofu}, S\BPBI I.%
\BCBL {}\ \BBA {} {Smith}, E\BPBI J.%
\end{APACrefauthors}%
\unskip\
\newblock
\APACrefYearMonthDay{1988}{{\APACmonth{08}}}{}.
\newblock
{\BBOQ}\APACrefatitle {{Origin of interplanetary southward magnetic fields
  responsible for major magnetic storms near solar maximum (1978-1979)}}
  {{Origin of interplanetary southward magnetic fields responsible for major
  magnetic storms near solar maximum (1978-1979)}}.{\BBCQ}
\newblock
\APACjournalVolNumPages{Journal of Geophysical Resarch}{93}{A8}{8519-8531}.
\newblock
\begin{APACrefDOI} \doi{10.1029/JA093iA08p08519} \end{APACrefDOI}
\PrintBackRefs{\CurrentBib}

\bibitem [\protect \citeauthoryear {%
{Vanninathan}%
, {Veronig}%
, {Dissauer}%
\BCBL {}\ \BBA {} {Temmer}%
}{%
{Vanninathan}%
\ \protect \BOthers {.}}{%
{\protect \APACyear {2018}}%
}]{%
vanninathan:2018}
\APACinsertmetastar {%
vanninathan:2018}%
\begin{APACrefauthors}%
{Vanninathan}, K.%
, {Veronig}, A\BPBI M.%
, {Dissauer}, K.%
\BCBL {}\ \BBA {} {Temmer}, M.%
\end{APACrefauthors}%
\unskip\
\newblock
\APACrefYearMonthDay{2018}{{\APACmonth{04}}}{}.
\newblock
{\BBOQ}\APACrefatitle {{Plasma Diagnostics of Coronal Dimming Events}} {{Plasma
  Diagnostics of Coronal Dimming Events}}.{\BBCQ}
\newblock
\APACjournalVolNumPages{The Astrophysical Journal}{857}{1}{62}.
\newblock
\begin{APACrefDOI} \doi{10.3847/1538-4357/aab09a} \end{APACrefDOI}
\PrintBackRefs{\CurrentBib}

\bibitem [\protect \citeauthoryear {%
{Vourlidas}%
, {Lynch}%
, {Howard}%
\BCBL {}\ \BBA {} {Li}%
}{%
{Vourlidas}%
\ \protect \BOthers {.}}{%
{\protect \APACyear {2013}}%
}]{%
vourlidas:2013}
\APACinsertmetastar {%
vourlidas:2013}%
\begin{APACrefauthors}%
{Vourlidas}, A.%
, {Lynch}, B\BPBI J.%
, {Howard}, R\BPBI A.%
\BCBL {}\ \BBA {} {Li}, Y.%
\end{APACrefauthors}%
\unskip\
\newblock
\APACrefYearMonthDay{2013}{{\APACmonth{05}}}{}.
\newblock
{\BBOQ}\APACrefatitle {{How Many CMEs Have Flux Ropes? Deciphering the
  Signatures of Shocks, Flux Ropes, and Prominences in Coronagraph Observations
  of CMEs}} {{How Many CMEs Have Flux Ropes? Deciphering the Signatures of
  Shocks, Flux Ropes, and Prominences in Coronagraph Observations of
  CMEs}}.{\BBCQ}
\newblock
\APACjournalVolNumPages{\solphys}{284}{1}{179-201}.
\newblock
\begin{APACrefDOI} \doi{10.1007/s11207-012-0084-8} \end{APACrefDOI}
\PrintBackRefs{\CurrentBib}

\bibitem [\protect \citeauthoryear {%
{Vr{\v{s}}nak}%
}{%
{Vr{\v{s}}nak}%
}{%
{\protect \APACyear {2001}}%
{\protect \APACexlab {{\protect \BCnt {1}}}}}]{%
vrsnak:2001}
\APACinsertmetastar {%
vrsnak:2001}%
\begin{APACrefauthors}%
{Vr{\v{s}}nak}, B.%
\end{APACrefauthors}%
\unskip\
\newblock
\APACrefYearMonthDay{2001{\protect \BCnt {1}}}{{\APACmonth{08}}}{}.
\newblock
{\BBOQ}\APACrefatitle {{Deceleration of Coronal Mass Ejections}} {{Deceleration
  of Coronal Mass Ejections}}.{\BBCQ}
\newblock
\APACjournalVolNumPages{Solar Physics}{202}{1}{173-189}.
\newblock
\begin{APACrefDOI} \doi{10.1023/A:1011833114104} \end{APACrefDOI}
\PrintBackRefs{\CurrentBib}

\bibitem [\protect \citeauthoryear {%
{Vr{\v{s}}nak}%
}{%
{Vr{\v{s}}nak}%
}{%
{\protect \APACyear {2001}}%
{\protect \APACexlab {{\protect \BCnt {2}}}}}]{%
vrsnak:2001b}
\APACinsertmetastar {%
vrsnak:2001b}%
\begin{APACrefauthors}%
{Vr{\v{s}}nak}, B.%
\end{APACrefauthors}%
\unskip\
\newblock
\APACrefYearMonthDay{2001{\protect \BCnt {2}}}{{\APACmonth{11}}}{}.
\newblock
{\BBOQ}\APACrefatitle {{Dynamics of solar coronal eruptions}} {{Dynamics of
  solar coronal eruptions}}.{\BBCQ}
\newblock
\APACjournalVolNumPages{\jgr}{106}{A11}{25249-25260}.
\newblock
\begin{APACrefDOI} \doi{10.1029/2000JA004007} \end{APACrefDOI}
\PrintBackRefs{\CurrentBib}

\bibitem [\protect \citeauthoryear {%
{Vr{\v{s}}nak}%
\ \protect \BOthers {.}}{%
{Vr{\v{s}}nak}%
\ \protect \BOthers {.}}{%
{\protect \APACyear {2010}}%
}]{%
vrsnak:2010}
\APACinsertmetastar {%
vrsnak:2010}%
\begin{APACrefauthors}%
{Vr{\v{s}}nak}, B.%
, {{\v{Z}}ic}, T.%
, {Falkenberg}, T\BPBI V.%
, {M{\"o}stl}, C.%
, {Vennerstrom}, S.%
\BCBL {}\ \BBA {} {Vrbanec}, D.%
\end{APACrefauthors}%
\unskip\
\newblock
\APACrefYearMonthDay{2010}{{\APACmonth{03}}}{}.
\newblock
{\BBOQ}\APACrefatitle {{The role of aerodynamic drag in propagation of
  interplanetary coronal mass ejections}} {{The role of aerodynamic drag in
  propagation of interplanetary coronal mass ejections}}.{\BBCQ}
\newblock
\APACjournalVolNumPages{\aap}{512}{}{A43}.
\newblock
\begin{APACrefDOI} \doi{10.1051/0004-6361/200913482} \end{APACrefDOI}
\PrintBackRefs{\CurrentBib}

\bibitem [\protect \citeauthoryear {%
{Wenzel}%
, {Marsden}%
, {Page}%
\BCBL {}\ \BBA {} {Smith}%
}{%
{Wenzel}%
\ \protect \BOthers {.}}{%
{\protect \APACyear {1992}}%
}]{%
wenzel:1992}
\APACinsertmetastar {%
wenzel:1992}%
\begin{APACrefauthors}%
{Wenzel}, K\BPBI P.%
, {Marsden}, R\BPBI G.%
, {Page}, D\BPBI E.%
\BCBL {}\ \BBA {} {Smith}, E\BPBI J.%
\end{APACrefauthors}%
\unskip\
\newblock
\APACrefYearMonthDay{1992}{{\APACmonth{01}}}{}.
\newblock
{\BBOQ}\APACrefatitle {{The ULYSSES Mission}} {{The ULYSSES Mission}}.{\BBCQ}
\newblock
\APACjournalVolNumPages{Astronomy and Astrophysics Supplement
  Series}{92}{}{207}.
\PrintBackRefs{\CurrentBib}

\bibitem [\protect \citeauthoryear {%
{Wu}%
, {Liou}%
, {Hutting}%
\BCBL {}\ \BBA {} {Wood}%
}{%
{Wu}%
\ \protect \BOthers {.}}{%
{\protect \APACyear {2022}}%
}]{%
wu:2022}
\APACinsertmetastar {%
wu:2022}%
\begin{APACrefauthors}%
{Wu}, C\BHBI C.%
, {Liou}, K.%
, {Hutting}, L.%
\BCBL {}\ \BBA {} {Wood}, B\BPBI E.%
\end{APACrefauthors}%
\unskip\
\newblock
\APACrefYearMonthDay{2022}{{\APACmonth{08}}}{}.
\newblock
{\BBOQ}\APACrefatitle {{Magnetohydrodynamic Simulation of Multiple Coronal Mass
  Ejections: An Effect of ``Pre-events''}} {{Magnetohydrodynamic Simulation of
  Multiple Coronal Mass Ejections: An Effect of ``Pre-events''}}.{\BBCQ}
\newblock
\APACjournalVolNumPages{\apj}{935}{2}{67}.
\newblock
\begin{APACrefDOI} \doi{10.3847/1538-4357/ac7f2a} \end{APACrefDOI}
\PrintBackRefs{\CurrentBib}

\end{thebibliography}
\end{document}